\documentclass[preprint,aps,nofootinbib,showpacs,showkeys,amsmath,amssymb]{revtex4}
\usepackage{graphicx}
\newcommand{\bR}{{\mathbb R}}

\newcommand{\bC}{{\mathbb C}}
\newcommand{\go}{\mathfrak}                 % gothic font (usual)
\newcommand{\ga}{\go A}
\newcommand{\gH}{\go H}
\newcommand{\gHa}{{\go H}_\alpha}
\newcommand{\ha}{H_\alpha}
\newcommand{\aaa}{A_\alpha}
\newcommand{\gba}{g_{\beta\alpha}}
\newcommand{\la}{\Lambda_\alpha}
\newcommand{\be}{\begin{equation}}
\newcommand{\ee}{\end{equation}}
\newcommand{\n}{{\bf n}}
\newcommand{\rr}{{\bf r}}
\newcommand{\vni}{\vec{n}_i}

   \newcommand*{\bracket}[1]{\langle#1\rangle}

   \newcommand*{\De}{\Delta}

\newcommand{\gab}{g_{\alpha\beta}}
\newtheorem{theorem}{Theorem}
\DeclareGraphicsExtensions{.jpg,.pdf,.gif,.png,.eps}
\begin{document}
\title{Simultaneous Measurement of Non-commuting Observables \\ and Quantum
Fractals on Complex Projective Spaces\footnote{Paper based on the talk
given at the  Rencontre G\'eom\'etrie et Physique, CIRM, Marseille,
October 13--17, 2003. Dedicated to the memory of Moshe Flato.}}
\author{Arkadiusz \surname{Jadczyk}}
\affiliation{International Institute of Mathematical Physics,\\
7516 Paris, France} \email{ajad@quantumfuture.net}
\homepage{http://quantumfuture.net/quantum_future/}
\date{\today}
\begin{abstract} The simultaneous measurement of several noncommuting
observables is modeled by using semigroups of completely positive
maps on an algebra with a non-trivial center. The resulting
piecewise-deterministic dynamics leads to chaos and to nonlinear iterated
function systems (quantum fractals) on complex projective spaces.
\end{abstract}
\maketitle
\section{Introduction}
From the very beginning quantum mechanics has been formulated in
rather abstract mathematical terms: operators, commutators,
eigenvalues, eigenvectors, etc. For the most part, the
accompanying physical interpretations were discovered as surprises
rather than due to any deeper understanding of what all this new
theory was about. Much of the axiomatization of quantum theory
originated in the works of John von Neumann, culminating in his
classic monograph ``Mathematical Foundations of Quantum Theory''
\cite{Neu}. But physics is not always as simple as mathematicians
would like it to be. Even if the criteria of mathematical elegance
and simplicity are often useful in sorting out candidates for
possible formal descriptions of reality, Nature herself has proven
to have a sense of elegance that quite often goes deeper than what
we would naively expect. The unfortunate result of the lack of
deeper understanding of the physical foundations of quantum theory
(as exemplified by the famous discussions between Einstein and
Bohr, with Einstein exclamating: ``God does not play dice'', and
Bohr responding: ``Einstein, stop telling God what to do'') was
that the theory has been axiomatized, including the concept of
`measurement'. In this way for many many years only a few brave
physicists dared to notice that the emperor has no clothes and say
it aloud.  As we have stressed elsewhere \cite{jad94c} John Bell
\cite{bell89,bell90} deplored the misleading use of the term
`measurement' in quantum theory.\footnote{``Why did such serious
people take so seriously axioms which now seem so arbitrary? I
suspect that they were misled by the pernicious misuse of the word
`measurement' in contemporary theory.'' writes John Bell in
\cite{bell87a}} He opted for banning this word altogether from our
quantum vocabulary, together with other vague terms such as
`macroscopic', `microscopic', and `observable'. (Today he would
probably add to his list two other terms of similarly dubious
validity: `environment', and `environmentally induced
decoherence'.) He suggested that we ought to replace the term
`measurement'
 with that of `experiment' \cite{bell87a}, and
also not to even speak of `observables' (the things that seem to
call for an `observer') but to introduce, instead, the concept of
`beables' \cite{bell87b} --~the things that objectively
`happen--to--be (or not--to--be)', independent of whether there is
some `observer', even if only in the future \cite{wheeler83}, or
not. In his scrupulous critical analysis of the quantum
measurement problem \cite{bell90}, ``Against Measurement,'' John
Bell indicates that to make sense of the usual mumbo jumbo one
must assume either that (i) in addition to the wave function
$\psi$ of a system one must also have variables describing the
classical configuration of the apparatus or (ii) one must abrogate
the Schr\"odinger evolution during measurement, replacing it by
some sort of collapse dynamics.

The theory of quantum events (EQT)\footnote{This theory is also
known as EEQT - `Event Enhanced Quantum Theory'. In the present
paper we consistently replaced EEQT with EQT (except in the
references) - which is more convenient.}, outlined in Section 2,
combines (i) and (ii): there are additional classical variables,
commonly referred to as `superselection rules', and because of the
coupling between these variables and the quantum degrees of
freedom, the evolution is not exactly the unitary Schr\"odinger
evolution, and it leads to collapses, in particular in
measurement-like situations.

It is to be noted that Bell criticized both (i) and (ii), because
both ascribe a special fundamental role to `measurement', which
seems implausible and makes vagueness unavoidable. EQT takes his
valid criticism into account. In EQT we make a distinction between
a measurement and an experiment. Both have a definite meaning
within EQT. According to the general philosophy of EQT, our
universe, one that we perceive and are trying to describe and
understand, can be considered as being `an experiment' --
performed by Nature herself. This is in total agreement with Bell;
it is also in agreement with the philosophy of John Wheeler, as
outlined in \cite{wheeler83,wheeler89}. John A. Wheeler stressed
repeatedly \cite{wheeler83}: ``No elementary quantum phenomenon is
a phenomenon until it is a registered (`observed,' `indelibly
recorded') phenomenon.''  But, he did not give a definition of
`being recorded' (though he stressed that human `observers' are
neither primary nor even necessary means by which quantum
potentials become `real') -- and we now understand why: Because
such a definition could not have been given within the orthodox
quantum theory. It is given in EQT -- see Section $2$ below.

Historically, physicists arrived at the quantum formalism by a
formal process known as `quantization'. Bohr's quantization,
Sommerfeld's quantization, geometric quantization, deformation
quantization ... Today there is a multitude of formal quantization
procedures, each leading to the end result that classical
quantities are being formally replaced by linear operators that,
in general, do not commute. The same components of position and
momentum do not commute. Different components of spin do not
commute. In each case the quantum commutation relations involve
Planck's constant on the right hand side. It is normally
considered that it is not possible to measure simultaneously
several noncommuting observables. One usually quotes in this
respect the celebrated Heisenberg's uncertainty relations. One
must notice that, in his classic monograph \cite{Neu}, John von
Neumann was very careful in this respect, and he stressed
explicitly that formal mathematical relations in no way indicate
impossibility of a simultaneous and precise measurements of, say,
position and momentum. He relied completely, in his account of the
`physical interpretation' of uncertainty relations, on the
`thought experiments' of Bohr and Heisenberg. Various textbook
authors treat the subject in a different way. A reasonable and
modern account of the problem is presented \cite{ingarden89},
where the authors present the standard derivation of Robertson's
inequality (\ref{eq:Robertson}), and then add the following
commentary:
\begin{quotation}
``It follows from the Heisenberg's uncertainty principle, and from the Theorem VII.1,
that momentum and position are not commensurable, that is there is no generalized observable A such that
\begin{eqnarray}
A(\Delta\times{\bf R}^1) &= E^Q(\Delta),\nonumber\cr A({\bf
R}^1\times\Delta) &= E^P(\Delta),\nonumber\cr
\end{eqnarray}
for $\Delta\subset {\cal B}({\bf R}^1)$. However, that does not
mean that quantum mechanics excludes the possibility of a
simultaneous measurement of $P$ and $Q$. In experimental technique
we are dealing with a simultaneous measurement of the momentum and
position. For instance, we observe a particle in a Wilson chamber.
From the observation of a particle track we determine its momentum
and position. For a charged particle we deduce its momentum by
placing the Wilson chamber in a magnetic field, and by measuring
the curvature of the track. Even in a situation when we are only
measuring the momentum of the particle, we have some knowledge of
its position, for instance that the particle is within the volume
of the measuring apparatus. The point is that in those situations
we are not talking about the simultaneous measurement in the exact
sense (description by spectral measures), but only about an
approximate measurement, with a given uncertainty - such as a
measurement described in example 6, section 12.1. The advantage of
the formalism of generalized observables [{\it i.e. using positive
operators rather than idempotents\ }] is a possibility of a
mathematical description of such a situation.''
\end{quotation}
In EQT indeed we are using positive operators and  projections,
but that is not important for the very modeling of the
simultaneous measurement of non-commuting observables. In EQT
fuzziness results in self-similarity and fractal patterns, but is
not a necessary feature of the chaotic dynamics resulting from
noncomeasurabilty. Masanao Ozawa, in a recent series of papers
\cite{ozawa2001,ozawa2002,ozawa2003a,ozawa2003b,ozawa2003c},
reviewed the actual status of theories of state reduction and
joint measurement of non-commuting observables. Let us recall that
for any pair of observables $A$ and $B$ we have the following
relation \cite{Rob29}:
\begin{equation}\label{eq:Robertson}
\De_\rho A\De_\rho B\ge \frac{1}{2}|\bracket{[A,B]}_\rho|,
\end{equation}
where $\bracket{\cdots}_\rho$ stands for the mean value in the
given state $\rho$, $\De_\rho A$ and $\De_\rho B$ are the standard
deviations of $A$ and $B$, defined by $\De_\rho
X=(\bracket{X^{2}}_\rho-\bracket{X}^{2}_\rho)^{1/2}$ for $X=A,B$,
and the square bracket stands for the commutator, i.e.,
$[A,B]=AB-BA$. In particular, for two conjugate observables $Q$
and $P$, which satisfy the canonical commutation relation
\begin{equation}\label{eq:CCR}
[Q,P]=i\hbar,
\end{equation}
we obtain Kennard's inequality \cite{Ken27}
\begin{equation}\label{eq:HUR}
\De_\rho Q \De_\rho P\ge\frac{\hbar}{2}.
\end{equation}
In \cite{ozawa2001} Ozawa concludes that \begin{quotation}``...
the prevailing Heisenberg's lower bound for the noise-disturbance
product is valid for measurements with independent intervention,
but can be circumvented by a measurement with dependent
intervention. An experimental confirmation of the violation of
Heisenberg's lower bound is proposed for a measurement of optical
quadrature with currently available techniques in quantum
optics.''
\end{quotation} In a recent paper of this series
\cite{ozawa2003b} Ozawa writes \begin{quotation} ``Robertson's and
Kennard's relations are naturally interpreted as the limitation of
state preparations or the limitation of the ideal independent
measurements on identically prepared systems \cite{Bal70,Per93}.
Moreover, the standard deviation, a notion dependent on the state
of the system but independent of the apparatus, cannot be
identified with the imprecision of the apparatus such as the
resolution power of the $\gamma$ ray microscope. Thus, it is still
missing to correctly describe the unavoidable imprecisions
inherent to joint measurements of noncommuting observables.
''\end{quotation}

Although our criticism of the standard treatment of the
measurement process and of the interpretation of the uncertainty
relations goes much deeper, we do agree with the above
conclusions.
\subsection{Quantum Events Theory - Duality}
EQT starts with the realization that any formal description of
Reality must have a dual, partly classical and partly quantum
nature. Those who deny this, contradict themselves by the very act
of denying. Indeed, as stressed already by Niels Bohr, the
sentences that they write, the conclusions they come to, are all
classical in nature. In \cite{bell87a} John Bell writes:
\begin{quotation}
``But we cannot include the whole world in the wavy part. For the
wave of the world is no more like the world we know than the
extended wave of the single electron is like the tiny flash on the
screen. We must always exclude part of the world from the wavy
`system', to be described in a `classical' `particulate' way, as
involving definite events rather than just wavy possibilities.''
\end{quotation}
The fact of communicating anything through
some channel, in finite time, is an `event' -- and as such, it is
classical. It happens. However, there are no events in standard
quantum theory, they do not belong to quantum dynamics, and the
standard quantum theory does not provide us with any understanding
of why, how, and when they happen. That is why the standard theory is
incomplete. In 1986 John Bell, envisioning a possibility of creating a
new, more complete theory wrote \cite{bell86}:
\begin{quotation}
``And surely in fundamental theory this merging [{\it of classical
and quantum}\ ] should be described not just by vague words but by
precise mathematics? This mathematics would allow electrons to
enjoy the cloudiness of waves, while allowing tables and chairs,
and ourselves, and black marks on photographs, to be rather
definitely in one place rather than another, and to be described
in `classical' terms. The necessary technical theoretical
development involves introducing what is called `nonlinearity',
and perhaps what is called `stochasticity', into the basic
`Schr{\"o}dinger equation'.''
\end{quotation}
EQT is a step in this direction, a step involving nonlinearity,
non-unitarity, and stochasticity. The new mathematics of EQT,
based on piecewise deterministic processes,  enables us also to
understand why the simultaneous measurement of noncommuting
observables leads to chaotic dynamics that could not have been
anticipated by the founders of quantum theory.
\subsection{Central classical observables}
In EQT we assume that, for one reason or another, the important
object is a $\star$-algebra of operators $\ga .$\footnote{All
algebras and all Hilbert spaces discussed here are over the field
of complex numbers $\bC.$} For historical reasons $\ga$ is called
an `algebra of observables', even if only normal operators, that
is those which commute with their adjoints, are believed to be
directly related to observable physical quantities. In EQT the
elements of $\ga,$ even if they can represent `physical
quantities', can neither be observed nor do they represent, as it
is assumed within the standard interpretation `observational
procedures' -- except in a limit that is rather unrealistic. We
will see that operators in $\ga$ do exactly what they are supposed
to do: they {\sl operate}\ on states to produce new states that
result from quantum events. They implement quantum jumps that
accompany any event and any information gain related to the
quantum system. It should be noted that in EQT we do not import
any a priori probabilistic interpretation of the standard quantum
theory. All interpretation is being derived from the Piecewise
Deterministic Process (PDP) described below. Interpretation of
eigenvectors, eigenvalues, mean values of observables, etc. should
be derived from the dynamics of EQT. Part of the standard wisdom
about eigenvalues and eigenvectors can, in fact, be justified
within EQT, and so we will use it as a heuristic tool for
constructing mathematical models of `real world' situations.
 The algebra $\ga$ is usually assumed to be a $C^\star$ or a von
Neumann algebra, but EQT can work also in spaces with indefinite
scalar product or within a Clifford algebra framework. A generic
algebra $\ga$ will have a nontrivial center ${\go Z}$ -- the set
of all $A\in {\go A}$ which commute with all the elements of ${\go
A}.$ In particular ${\go Z}$ is Abelian -- it represents the
classical subsystem. Algebras with trivial center (i.e. center
consisting of operators that are complex multiples of the
identity) are called {\em factors}. Physicists insisting on the
idea that there are no genuine classical degrees of freedom are,
in fact, insisting on the idea that only factors should be used
for an algebraic description of quantum systems. While it is true
that every algebra can be decomposed, essentially uniquely, into a
direct sum (or integral) of factors, restricting to factors alone
is like restricting to prime numbers alone. While it is true that
any integer can be decomposed into a product of prime numbers,
insisting on the idea that only prime numbers should be used would
be simply silly. Atoms build molecules. There would be no life
without molecules. Similarly factors build more complex
non-factors. According to our definition below, there would be no
`events' without non-factors! Thus there would be no data
(recording a datum is an event) that could be used in experiments.

Each Abelian algebra has only one-dimensional irreducible
representations. These are called characters, and the set of all
characters of ${\go Z}$ is called the spectrum of ${\go Z}.$ By
quite general representation theorems, each Abelian algebra is
naturally isomorphic to an algebra of functions over its spectrum
(continuous, measurable etc., depending on the type of the
algebra). For simplicity we will assume that the spectrum of ${\go
Z}$ is discrete -- countable, or even finite. With proper care we
could consider more general cases -- as for instance in the
SQUID-tank model, where the spectrum of ${\go Z}$ is a symplectic
manifold - the phase space of a radio-frequency oscillator (cf.
\cite{blaja93c,olk97}, and also \cite{olk99,blaolk99a,blaolk99b}
for other examples of working EQT models with a continuous
spectrum of ${\go Z}$). Heuristically the points of the spectrum
of ${\go Z}$ are the `pointer positions' --~that is, states of the
classical subsystem~-- we will denote the spectrum of ${\go Z}$ by
the letter ${\cal C}.$ Discrete changes of states of ${\cal C}$
are called {\em events}\ . When the set of classical states is
discrete, then any change of it is discrete. But, for instance, in
models with a continuous spectrum (as, for instance, when ${\cal
C}$ is a phase space $\{q,p\}$) we will have a continuous
evolution of the state of ${\cal C}$ that is interrupted by
events, for instance jumps in the momentum $p$ (instantaneous
boosts) in ${\cal C}.$

\section{An Outline of the formal scheme of Quantum Events Theory (EQT)}
\subsection{Completely Positive Maps}\
Historically, EQT started with an attempt at describing time
evolution of a system with a non-trivial center, in the simplest
case with ${\go A}={\go A}_q\otimes{\go A}_{cl},$ and ${\go
Z}\approx {\go A}_{cl},$ where there would be a dynamical coupling
and mutual exchange of information between the quantum and the
classical degrees of freedom. Because algebra automorphisms
preserve the center of any algebra, it was clear that
automorphisms could not be used to this end. In a private
communication with the author, Rudolph Haag, long ago, expressed
his doubts as to the physical significance of the algebraic
product in the algebra of observables. Even if the product $AB$ is
useful in setting up the canonical commutation relations, the
product of observables is not itself an observable and, therefore,
need not be necessarily preserved by time evolution when
irreversible recording is taking place. What seems to have
physical meaning is positivity in the algebra, therefore the
simplest generalization of the automorphic evolution takes us to
semigroups of positive maps. Positivity itself is not a stable
condition. Adding spurious degrees of freedom which do not
participate in the dynamics can destroy positivity. The more
stable condition is called `complete positivity'. It is defined as
follows:

Let ${\go A},{\go B}$ be $C^{\star}$--algebras.  A linear map $\phi :  {\go A}
\rightarrow {\go B}$ is {\sl Hermitian} if $\phi  (A^{\star}) = \phi
 (A)^{\star }. $\ It is {\sl positive} iff $A\geq 0 , $ $A\in {\go A}$ implies $
\phi  (A)\geq 0 . $ Because Hermitian elements of a
$C^{\star}$--algebra are differences of two positive ones -- each
positive map is automatically Hermitian.  Let ${\go M}_n$ denote
the $n$ by $n$ matrix algebra,  and let $ {\go M}_n ({\go A}) =
{\go M}_n \otimes {\go A}$ be the algebra of $n\times n$ matrices
with entries from ${\go A}. $ Then ${\go M}_n ({\go A})$ carries a
natural structure of a $C^{\star}$--algebra. With respect to this
structure a  matrix ${\bf A}= (A_{ij})$ from ${\go M}_n  ({\go
A})$ is positive iff it is  a sum of matrices of the form $
(A_{ij}) =  (A_i^{\star } A_j ), \,  $ $A_i\in {\go A}.$ If ${\go
A}$ is an algebra of operators on a Hilbert space ${\go H}$,  then
${\go M}_n ({\go A})$ can be considered as acting on ${\go H}^n
\doteq {\go H} \otimes \bC^n = \oplus_{i=1}^n {\go H} . $
Positivity of ${\bf A}= (A_{ij})$ is then equivalent to $ ({\bf
\Psi}, {\bf A} {\bf \Psi} )\geq 0\,  , \,  {\bf \Psi} \in {\go
H}^n ,  $ or equivalently,  to $\sum_{i, j}  (\Psi_i ,  A_{ij}
\Psi_j ) \geq 0 $ for all $\Psi_1, \ldots , \Psi_n \in {\go H} . $

A positive map $ \phi$ is said to be {\sl completely positive} or,
briefly,   CP
 iff $\phi \otimes id_n
: {\go A }\otimes {\go M}_n \rightarrow {\go B}\otimes {\go M}_n $ defined
by $  (\phi\otimes id_n)  (A\otimes M ) = \phi  (A)\otimes M , \,  M\in {\go M}_n
$, is positive for all $n=2, 3, \ldots . $ When written explicitly,  complete
positivity is equivalent to \be \sum_{i, j=1}^n B_i^{\star}\phi
 (A_i^{\star}A_j)B_j \geq 0 \ee for every $A_1, \ldots , A_n \in {\go A}$ and
$B_1, \ldots , B_n \in {\go B} . $ In particular every homomorphism of
$C^{\star}$ algebras is completely positive.  One can also show that if either
${\go A}$ or ${\go B}$ is Abelian,  then positivity implies complete
positivity.  Another important example:  if ${\go A}$ is a $C^{\star }$
algebra of
operators on a Hilbert space ${\go H}$,  and if $V\in {\go B} ({\go H}) , $
then $\phi  (A) = VAV^{\star}$ is a CP map $\phi :  {\go A}
\rightarrow \phi ({\go A}) . $

 In the quantum dynamics of open systems the unitary time evolution described by the Schr\"odinger
   equation is usually replaced by a semigroup of completely positive
   maps (also known as a `dynamical semigroup') \cite{alicki87,alicki02}.
   Usually such semigroups are being studied on the von Neumann algebra
   ${\go A}$ of all bounded linear operators ${\ga}=L (\go H )$ on a separable
   Hilbert space ${\go H}.$ In the algebraic framework \cite{emch} we learn
   that more general von Neumann algebras can also appear in physical
   applications, in particular, as discussed above, algebras with a nontrivial center
   ${\go Z}=\ga\cap\ga^\prime ,$ where $\ga^\prime$ is the
   commutant of $\ga .$ The nontrivial central elements lead to
   superselection sectors (cf. \cite{landsman91}, and references therein), and,
   due to their commutativity with all observables, they represent the `classical observables' of the
   theory. Applying open system dynamics to an algebra with a nontrivial center brings in new
   possibilities, with an interesting new result that there is a
   one-to-one correspondence between a class of completely positive semigroups
   and piecewise deterministic random processes (PDP -- cf. \cite{davis93}) on the space
   of pure states of the algebra.  It has been shown that, in
some cases, the associated piecewise deterministic process can be
interpreted as a nonlinear iterated function system (IFS) on a complex
projective space of rays in
the Hilbert space ${\go H}$, with a
fractal attractor, and with a range of Lyapunov's exponents depending on
a particular value of the coupling constant in the semigroup
generator \cite{blajaol99a}

In the present paper the algebra $\ga$ of observables will be
assumed to be a von Neumann algebra. The points of the spectrum of
its center ${\go Z}$ represent (pure) states of the Abelian
subalgebra (superselection sectors). We will denote these states
$\alpha=1,\ldots ,m$. The algebra $\ga$ is then of the form $\ga =
\oplus_{\alpha=1}^m \ga_\alpha $, where $\ga_\alpha$ are factors
(that is they have a trivial center). We will be interested in the
simplest case, where $\ga_\alpha =L({\go H}_\alpha)$, where $\gHa$
is a Hilbert space of dimension (possibly infinite) $n_\alpha .$
Thus every element $A\in \ga$ is represented by a family
$\{A_\alpha\}$ of operators $A_\alpha \in  L(\gHa )$, or as a
block diagonal matrix operator $A = diag (A_1,\ldots ,A_m )$ on
$\gH\doteq\oplus_{\alpha=1}^m \gHa .$ Every normal state $\rho$ of
$\ga$ is represented by a density matrix on $\gH$ , that is by a
family $\{\rho_\alpha\}$ of positive, trace-class operators on
$\gHa$, with $\sum_{\alpha = 1}^m Tr (\rho_\alpha )=1,$ and
$\rho(A)=\sum_{\alpha =1}^m Tr(\rho_\alpha A_\alpha ).$
\subsection{Dynamical Semigroups on an Algebra with a Center}
The most general form of a generator of a completely
positive semigroup is then given by the formula of Christensen and
Evans \cite{chr}, which generalizes the classical results of
Gorini, Kossakowski and Sudarshan \cite{koss} and of Lindblad
\cite{lin} to the case of an arbitrary $C^{\star }$--algebra.
It is worthwhile to cite, after Lindblad, his original
motivation:
\begin{quotation} The dynamics of a finite closed quantum system is
conventionally represented by a one--parameter group of unitary
transformations in Hilbert space.  This formalism makes it difficult to
describe irreversible processes like the decay of unstable particles,
approach to thermodynamic equilibrium and measurement processes [$\ldots $].
It seems that the only possibility of introducing an irreversible behavior
in a finite system is to avoid the unitary time development altogether
by considering non--Hamiltonian systems.\end{quotation}
\begin{theorem}[Christensen -- Evans]
Let $\alpha_t = \exp  (L t)$ be a norm--continuous semigroup of CP
maps of a $C^{\star }$-- algebra of operators ${\go A}\subset
L({\go H}) . $ Then there exists a CP map $\phi$ of ${\go A}$ into
the ultraweak closure ${\bar {\go A}}$ and an operator $K\in {\bar
{\go A}}$ such that the generator $L$ is of the form: \be L (A) =
\phi  (A) + K^{\star }A + AK \,  . \ee
\end{theorem}
The set of all CP maps $\phi: \ga\rightarrow \ga$ is convex. Of
particular interest to us are generators $L$ for which $\phi$ is
extremal. Arveson \cite{arv}, using the celebrated Stinespring
theorem \cite{stinespring}, proved that this is the case if and
only if  $\phi$ is of the form \be \phi  (A)=V^{\star }\pi  (A) V
\,  ,  \ee where $\pi$ is an irreducible representation of ${\ga}$
on a Hilbert space ${\go K} , $ and $V: {\cal H}\rightarrow {\go
K}$ is a bounded operator  (it must be,  however,  such that
$V^{\star}\pi ({\ga}) V \subset {\ga }$). Then $\phi (I)=V^{\star
}V.$ In the following we will assume that all $n_\alpha < \infty$,
then ${\bar {{\ga}}} = {\ga} , $ so that $K=\{K_\alpha\}\in {\ga}
. $ We will always assume that $\alpha_t (I) = I $ or,
equivalently, that $L (I)=0 . $  It is convenient to introduce $
H_\alpha=i (K_\alpha-K_\alpha^{\star })/2 \in L(\gHa), $ then from
$L (I)=0$ we get $K_\alpha+K_\alpha^{\star }=-\phi  (I)_\alpha , $
and so $K_\alpha=-iH_\alpha-\phi (1)_\alpha/2 . $ Therefore we
have \be L (A)_\alpha = i\left[H_\alpha, A_\alpha\right]+\phi
(A)_\alpha -\{ \phi (1)_\alpha , A_\alpha\}/2 ,  \ee where $\{\, ,
\, \}$ denotes the anticommutator. Using the Arveson result it is
easy to see that, in our case, $\phi$ is a non-zero extremal CP
map if and only if $V$ is if of the form $V=\{V_{\alpha\beta}\}$,
where only one matrix entry $V_{\alpha_0\beta_0}:
\gH_{\beta_0}\rightarrow \gH_{\alpha_0}$ is non-zero. Taking for
$\phi$ a sum of maps of such a type we end up with a generator $L$
of the form: \be L(A)_\alpha=i[\ha,\aaa]+\sum_\beta \gba^\star
A_\beta\ \gba - \frac{1}{2}\{\la,\aaa\},\label{eq:lioua} \ee where
$\gab \in L({\go H}_\beta,{\go H}_\alpha)$ and \be \la=\sum_\beta
\gba^\star \gba \in L(\gHa ). \ee
\subsection{The Liouville Equation for States}
Taking into account the duality between observables and states, given by the valuation $<\rho ,A>=Tr(\rho A)=\sum_{\alpha=1}^m Tr(\rho_\alpha A_\alpha ),$ the evolution equation
for the semigroup ${\dot A}= L(A)$ can be rewritten in terms of states:
\be
{\dot \rho}_\alpha=-i[H_\alpha,\rho_\alpha]+\sum_{\beta}
 \gab
\rho_\beta \gab^\star-\frac{1}{2}\{\Lambda_\alpha,\rho_\alpha\}.
\label{eq:lio} \ee Notice that the total trace is automatically
conserved:
$$\frac{d}{dt} Tr(\rho) = \sum_\alpha Tr({\dot \rho}_\alpha )=0.$$
In problems that are explicitly time-dependent, as it is in most
cases where there is an explicit intervention of the
`experimenter', who sets up the characteristics of the measuring
device according to the needs of the experiment, the maps $\phi$
and $K$, and thus the operators $H_\alpha$ and $\gab$ will depend
on time, and they will generate a family $\alpha_t$ of CP maps,
which will not have the semigroup property.
\subsection{Ensemble and Individual Descriptions\label{sec:ensemble}}
There are two descriptions in EQT: the ensemble description and
the individual description. The ensemble description is a
deterministic, smooth Liouville evolution of statistical states.
The individual description is piecewise deterministic process on
the space of pure states, where a continuous, nonunitary,
evolution is interrupted by discontinuous catastrophic events. One
goes from the individual to the ensemble description by averaging
over many sample paths. The averaging process smoothes out
discontinuouities and nonlinearities.

The jump probabilities in the process will be computed from the formula:

\be
p_{\alpha\rightarrow\beta}(\psi ,t)=\Vert g_{\beta\alpha}(t)\psi\Vert^2/<\psi , \Lambda_\alpha (t) \psi> .\label{eq:pab}
\ee

It has been shown in \cite{jakol95} that when the diagonal terms $g_{\alpha\alpha}$
all vanish, then there is a one-to-one correspondence between the solutions of the Liouville
equation (\ref{eq:lio}), and PDP processes on the space of pure states of the algebra $\ga$, where the
process realizing the solution of Eq. (\ref{eq:lio}) with the initial pure state $\rho=(0,\ldots ,|\psi_{\alpha_0} ><\psi_{\alpha_0}|,\ldots , 0)$ is described as follows: \\
\vskip 0.2cm \noindent {\bf PDP Process:} Given on input
$t_0,\alpha_0,$ and $\psi_0\in{\go H}_{\alpha_0}$, with $\Vert
\psi_0\Vert=1$, it produces on output $t_1,\alpha_1$ and
$\psi_1\in{\cal H}_{\alpha_1}$, with $\Vert\psi_1\Vert=1$. \vskip
0.2cm {\em
\noindent $1$) Choose uniform random number $r\in [0,1]$.\\
$2$) Propagate $\psi_0$ in ${\go H}_{\alpha_0}$ forward in time by
solving: \be {\dot \psi}(t)= \left(-iH_{\alpha_0}(t) -
\frac{1}{2}\Lambda_{\alpha_0}(t)\right){\psi}(t) \ee with initial
condition $\psi (t_0)=\psi_0$ until $t=t_1$, where $t_1$ is
defined by\footnote{Note that, as can be seen from the equation
(\ref{p}), the norm of $\psi(t)$ is a monotonically decreasing
function of $t$.} \be \Vert \psi(t_1) \Vert^2=r. \label{eq:time}
\ee
$3$) Choose a uniform random number $r^\prime\in [0,1].$\\
$4$) Run through the classical states $\alpha=1,2,\ldots ,m$ until
you reach $\alpha=\alpha_1$ for which \be \sum_{\alpha=1}^{\alpha_1}
p_{\alpha_0\rightarrow\alpha}(\psi(t_1),t_1)\geq r^\prime . \ee $5$) Set
$\psi_1=g_{\alpha_1\alpha_0}(t_1)\psi(t_1)/
\Vert g_{\alpha_1\alpha_0}(t_1)\psi(t_1) \Vert $.\\
} \vskip 0.2cm \noindent Time evolution of an individual system is
described by repeated application of the above algorithm, using
its output as the input for each next step. If we want to study
time evolution in a given interval $[t_{in},t_{fin}]$, then we
apply the algorithm by starting with $t_0=t_{in}$, repeating it
until we reach $t=t_{fin}$ somewhere in the middle of the
propagation in step 2). Then we normalize the resulting state.

According to the theory developed in Ref. \cite{davis93} the jump
process is an inhomogeneous Poisson process with intensity
function $\lambda_\alpha (t)=\left(\psi(t),\Lambda_\alpha
(t)\psi(t)\right)$. One way to simulate such a process is to move
forward in time by small time intervals $\Delta t$, and make
independent decisions for jumping with probability $\lambda_\alpha
(t) \Delta t$. This leads to the probability $p$ of a jump to
occur in the time interval $(t_0,t)$ given by \be p=1-\exp
(-\int_{t_0}^t \lambda_\alpha (s) ds).\label{p}\ee By using the
identity $\log f(t) -\log f (t_0) = \int_{t_0}^t {\dot f
(s)}/f(s)\; ds,$ with $f(s)\doteq ||\psi_\alpha (t)||^2,$ it is
easy to see that $p=1-\Vert \psi_\alpha(t) \Vert^2$ -- which
simplifies simulation -- as we did in the step 2) above. This
observation throws also some new light upon those approaches to
the quantum mechanical description of particle decays that were
based on non-unitary evolution.

By repeating the above event generating algorithm many times,
always starting with the same state at the same initial time
$t_0$, and ending it at the same final time $t,$ we will arrive at
different final states with different probabilities. Let
$\alpha_0,\psi_{\alpha_0},t_0$ be the initial state, and let
$\mu(\alpha_0,\psi_{\alpha_0},t_0;\alpha,\psi_\alpha,t)$ be the
probability density of arriving at the state
$(\alpha,\psi_\alpha)$ at time $t$. We may associate with this
probability distribution a family of density matrices:
\be\rho_\alpha (t) =  \int
\mu(\alpha_0,\psi_{\alpha_0},t_0;\alpha,\psi_\alpha,t)
|\psi_\alpha><\psi_\alpha | d \psi_\alpha , \ee so that
$\sum_\alpha {\mbox Tr} \rho_\alpha (t)=1$. This association is
many to one. We lose in this way information. Nevertheless, as
shown in \cite{jakol95}, the following theorem holds:
\begin{theorem}
The family $\rho_\alpha (t)$ satisfies the Liouville linear
differential equation (\ref{eq:lio}). Conversely, the PDP process
with values in the pure states $\alpha,\psi_\alpha$ described
above is the unique one leading to (\ref{eq:lio}).
\end{theorem}
The Liouville equation (\ref{eq:lio}) describes the time evolution
of the statistical states of the total system. This is the
standard, linear, master equation of statistical quantum physics,
an equation that describes infinite statistical ensembles, not
individual systems. Although the theorem quoted above tells us
that the event generating algorithm follows essentially uniquely
from the Liouville equation, we believe that it is the PDP process
rather than the statistical description that will lead to future
generalizations and extensions of the applicability of the quantum
theory.\footnote{Individual description gives us a deeper insight
into the real mechanism, and also is closer to reality, where some
experiments can be repeated only a few times, or even only once,
as it is with the Universe between Big Bang and Big Crunch.} For
instance, in the above formalism it is assumed that the operators
$g_{\alpha\beta}$ are linear. But they do not have to be. The
operators $\gab$ represent couplings between the quantum system
and a classical `detector pointer', and jumps represents `events'
i.e. changes of the pointer state. The formalism has been, in
particular, applied to the calculation of arrival times
\cite{blaja95f} and tunneling times for quantum particles
tunneling through a potential barrier \cite{palao,muga}, to the
calculation of relativistic time of arrival \cite{rus02a,rus02b},
and also for studying classical interventions in quantum systems
\cite{peres00b}.

\subsection{Simple Examples}
Physicists have long experience with constructing Hamiltonians
${H_\alpha}$ describing the action of external force fields and
different known interactions between particles. But how do we
construct the transition operators $g_{\alpha\beta}$? As has been
noticed by many authors, any `measurement' can be, in principle,
reduced to a position measurement. Once we know how to measure the
`pointer position', it is argued, it is enough to set up an
interaction between the apparatus and the system, both considered
as quantum systems, and, when the measurement is `done', read the
pointer position. While we do not think that life is that simple,
there is certainly some truth in the above, and therefore let us
start with a simple model of position measurement. The position
variable can be analyzed in terms of yes-no observations as to
whether a given region of space is occupied or not. Thus our first
example will describe a simple particle detector. In the next
section we will describe how a simultaneous monitoring of several
non-commuting observables can be modeled within EQT.
\subsubsection{A single detector \label{sg}}
A detector is a two-state device. It is often assumed that a
detector destroys the particle, but, as a typical track in a cloud
chamber shows, this need not be the case. There are several ways
of building a model of a detector, and we will describe the
simplest one, although not quite realistic. We would like to think
of a detector as a two-state device, with two meta-stable states,
denoted $0$ and $1$, able to jump from one state to another when
detecting a signal. We will assume zero relaxation time, so that
after detecting a signal, the detector is instantly ready to
detect another signal. Heuristically a particle passing close to
the detector can trigger its `flip' from $0$ to $1$, or from $1$
to $0.$ We will be interested only in the simplest case, when the
detection capability depends only on the particle location, and
not on its energy or other characteristics.\footnote{Adding a
relaxation time, even with an assigned probability distribution,
as well as modeling detectors with sensitivity dependent not only
on particle's location but also on its energy, or momentum, or
spin, is not a problem within EQT.}

Let us now specialize and consider a detector of particle presence
at a location $a$ in space (of $n$ dimensions). Our detector has a
certain range of detection and a certain efficiency. In a simple
model we encode these detector characteristics in a gaussian
function: \be g(x)=\kappa^{1/2} \left( \frac{1}{\sigma
\sqrt{\pi}}\right)^{n/2} \exp (-x^2/2\sigma^2),\label{eq:gaussian}
\ee where $\kappa$ is the detector sensitivity constant, $\sigma$
is a
width parameter, and $n$ stands for the number of space dimensions.\\
If the detector is moving in space along some trajectory $a(t)$,
and if the detector characteristics are constant in time and
space, then we put: $g_t (x)=g(x-a(t))$. Let us suppose that the
detector is in one of its states at $t=t_0$ and that the particle
wave function is $\psi_0(x)$. Then, according to the algorithm
described in section \ref{sec:ensemble}, the probability $p$ of
detection in the infinitesimal time interval $(t_0,t_0+\Delta t)$
is given by $p\approx\int g_{t_0}^2 (x) | \psi_0 (x) |^2 dx \cdot
\Delta t$. In the limit $\sigma\rightarrow 0$, when $g^2_t
(x)\rightarrow \kappa\, \delta (x-a(t))$ we get $p\approx\kappa
|\psi_0 (a(t_0)) |^2 \cdot \Delta t$. Thus, when $\Delta t <<
1/\kappa,$ we approximately recover the usual Born interpretation,
with the evident and necessary correction that the probability of
detection is proportional to the length of exposure
time of the detector. \\
\section{Measurement of
noncommuting observables}

EQT enhances the predictive power of the standard quantum theory,
and it does it in a rather simple way. Once enhanced it  predicts
new facts and straightens old mysteries. The model that we have
outlined above has several important advantages. One such
advantage is of a practical nature: for example in \cite{blaja93c}
it is shown how to generate pointer readings in a tank
radio--circuit coupled to a SQUID. In \cite{jad94b,jad94c} the
algorithm generating detection events of an arbitrary geometrical
configuration of particle position detectors, as for instance in a
Wilson chamber, has been derived. As a particular case, in a
continuous homogeneous limit we reproduced GRW spontaneous
localization model (cf \cite{jad94c} and references therein). Many
other examples come from quantum optics, since the Quantum Monte
Carlo model used there is a special case of our approach, namely
when {\em events} are not fed--back into the system and thus do
not really matter.

Another advantage of EQT is of a conceptual nature: in EQT we need
only one postulate: {\em that events can be observed}\,. All the
rest can and should be derived from this postulate. All
probabilistic interpretation, everything that we have learned, or
postulated, about eigenvalues, eigenvectors, transition
probabilities, etc. can be derived from the formalism of EQT. Thus
in \cite{blaja93a} we have shown that the probability distribution
of the eigenvalues of Hermitian observables can be derived from
the simplest, measurement-like, coupling. Moreover, in
\cite{jad94a} it was shown that EQT can also give definite
predictions for non--standard measurements, which are of
particular interest here, namely those involving noncommuting
observables, and that is so because in our scheme the
contributions $g_{\alpha\beta}$ from different, non-commuting,
devices {\em add} rather than multiply. In this respect, because
the measurement process is dynamical in our approach, it is like
adding non-commuting terms in a Hamiltonian -- nobody has
difficulty with adding a position function $A_\mu(x)$ to the
momentum $p_\mu$ in a Hamiltonian. They act simultaneously and
they act together.

Before we describe the model, and the resulting chaotic behavior
and strange attractor on the quantum state space, let us first
discuss in more detail the very question of the simultaneous
measurability of non-commuting observables. As we mentioned in the
Introduction, this subject has become quite controversial since
the early formulation of Heisenberg's uncertainty relations.
Mathematically these relations are precise and leave no doubt
about their validity, but the question of how to interpret them
physically and philosophically has become a subject of hot
discussions. Various views have been expressed on this subject.
Clearly there are different opinions. For instance one of the
early reviewers of the present paper wrote: ``QM, is a complete
theory, with precise, definite operational meanings attached to
the terms `measurement', `observable' \& c. Within its rules, the
simultaneous measurement of non-commuting observables is
impossible. This is well known and explained in any serious
textbook, which the author should be referred to.'' This sentence
evidently contradicts the other sentence, from the textbook by
Ingarden and Grabowski where, as already quoted in the
Introduction, the authors write: ``(...) However, that does not
mean that quantum mechanics excludes the possibility of a
simultaneous measurement of $P$ and $Q$. In experimental technique
we are dealing with a simultaneous measurement of the momentum and
position.'' It seems that the reviewer does not distinguish state
preparation procedure from measurements. To quote from Popper's
`Unended Quest' \cite{popper}:
\begin{quotation} ``The Heisenberg formula {\it do not refer to measurements};
which implies that the whole current `quantum theory of
measurement' is packed with misinterpretations. Measurements which
according to the usual interpretation of the Heisenberg formula
are `forbidden' are according to my results not only allowed, but
actually required for {\it testing} these very formula.''
\end{quotation}
 Hilary Putnam came to a similar conclusion \cite{putnam81}:
\begin{quotation} ``Recently I have observed that {\it it
follows from just the quantum mechanical criterion for measurement
itself}\ that the `minority view' is right to at least the
following extent: simultaneous measurements of incompatible
observables {\it can be made}\ . That such measurement cannot have
`predictive value' is true ...''
\end{quotation}
These words, written more than twenty years ago, suggested that
one should expect a chaotic behavior, and that this chaos and its
characteristics ought to be studied, both theoretically and
experimentally. Yet, for some reason, either no one noticed, or no
one got interested in looking into the problem quantitatively. Of
course the main theoretical obstacle was the unsolved
quantum-mechanical measurement problem. A good , critical,
discussion of current issues involved here, can be found in
\cite{landsman94}. In the abstract of his paper Landsman states:
\begin{quotation}``We attempt to clarify the main conceptual issues in approaches to
`objectification' or `measurement' in quantum mechanics which are
based on superselection rules. Such approaches  venture to derive
the emergence of classical `reality' relative to a class of
observers; those believing that the classical world exists
intrinsically and absolutely are advised against reading this
paper.''\end{quotation} Even if Landsman is guilty of using the
undefined, magical, word `environment', as, for instance, in
\begin{quotation}``The prototype approach (Hepp) where superselection sectors
are assumed in the state space of the apparatus is shown to be
untenable. Instead, one should couple system and  apparatus to an
environment, and postulate superselection rules for the latter,''
\end{quotation} he does a pretty good job in taking apart
different approaches and in analyzing their weaknesses.

\subsection{The simplest toy model - space and momentum are each only two-points.}
In this example we will describe the simplest possible toy model
of a simultaneous measurement of several non-commuting
observables. The most celebrated example is, of course, the
canonical pair of position and momentum observables. Although in
principle easy, technically it is difficult to simulate on a
computer, because the Hilbert space is infinite-dimensional. It is
also somewhat difficult to analyze analytically, due to its
continuous spectra. We will therefore choose here the maximally
simplified model - technically easy, almost trivial, and yet
demonstrating the whole idea.

The simplest, nontrivial `space' has just two points, we will
denote them `-1' and `1'. The `translation group' which operates
on these two points has two elements: the identity element and the
`flip' that exchanges these two points. We realize this simple
imprimitivity system\footnote{An imprimitivity system consists of
a spectral measure, and covariantly acting on this spectral
measure, a unitary representation of a group of transformations -
cf \cite{mackey,varadarajan}} in a two-dimensional Hilbert space
${\go H}=\bC^2.$ With the standard Pauli matrices
$\sigma_1,\,\sigma_2,\,\sigma_3$ defined by
$$\sigma_1\:=\:\left(\begin{array}{cc} 0& 1\\1& 0 \end{array}\right)\quad
\sigma_2\:=\:\left(\begin{array}{cc} 0& -i\\i& 0
\end{array}\right)\quad \sigma_3\:=\:\left(\begin{array}{cc} 1&
0\\0& -1 \end{array}\right)$$ we represent the `position operator'
by $\sigma_3$, and the `momentum operator' by $\sigma_1.$ Note
that in our case $\sigma_1$ represents the unitary `flip', while
$(I+\sigma_1)/2$ represents the `momentum'. We will need four
detectors, two for detecting the position eigenvalues $q=-1$ and
$q=+1$, and two for detecting the momentum eigenvalues $p=0$ and
$p=+1.$ As, formally, this is a particular case of a more general
situation, when we model a monitoring of several non-commuting
spin projections, in what follows we will discuss this more
general situation. As with the simple choice above, with four
detectors, we will consider a simple, highly symmetric geometric
pattern, so that the fractal and self-similarity effect are easily
recognizable. Our quantum system will be therefore a single spin
$1/2,$ with no spatial degrees of freedom. In order to construct a
model within the framework of EQT we need to specify the classical
system, its states, and the operations implementing transitions
between the states.  We will have a family of $n$ detectors, each
of them can be excited independently of the others, so that the
probability of two detectors being excited at the same time is
zero. Therefore a state of the classical system will be a sequence
of $n$ numbers, each number being $0$ or $1.$ There are $2^n$ of
such states, and a possible change of state consists of adding
$1\, \mbox{mod}\, 2$ at $i$-th place. For instance, if we have
three detectors, a possible transition between states can be
$\alpha=(1,0,1)\longrightarrow \beta=(0,0,1)$ - a flip of the
first detector. Only one detector can flip at a time. As for the
spin system, we will identify the Hilbert  spaces ${\go
H}_\alpha\equiv{\go H}\equiv{\bC}^2,\; \alpha=1,\ldots ,2^n ,$
corresponding to different states of the classical subsystems. In
this way the total algebra $\ga$ will be represented as a tensor
product $\ga=\ga_q \otimes \ga_c$ of its quantum and classical
parts. Because the quantum system is a two-state system, so the
quantum algebra $\ga_q=L({\go H})$ can be identified with the
algebra of $2\times 2$ complex matrices. Pure states of the spin
system are uniquely represented by points of the complex
projective space ${\bf P}_1 ({\bC})$, which is isomorphic to the
sphere ${\bf S}^2$ or, equivalently, by one-dimensional
projections of the form
$\frac{1}{2}(I\:+\:\vec{n}\cdot\vec{\sigma}),$ where $\vec{n}$ is
a unit vector in $\bR^3$ - pointing in the direction of the spin.

To be specific, let us consider a simple and symmetric
configuration of detectors, when the measuring apparatus consists
of six yes-no polarizers corresponding to $n=6$ spin directions
$\vni$, $i=\,1,...,n$, arranged at the vertices of a regular
octahedron along the directions $\vec{n}_i,\: i=1,\ldots ,n$:
$$\{\{\{0, 0, 1\}, \{1, 0, 0\}, \{0, 1, 0\},\\
   \{-1, 0, 0\}, \{0, -1, 0\}, \{0, 0, -1\}\}.$$
Notice that the six vectors sum up to zero
\begin{equation}\sum_{i=1}^{n}\vec{n}_i ={\bf 0}.\end{equation}
We may assume that our spin evolves according to the Hamiltonian
$H\,=\,\frac{\omega}{2}\sigma_3$, $\omega\geq 0.$ The coupling
between the spin system and the detectors is specified by choosing
six operators $a_i,$ which correspond to the six vectors $\vni$
\be
a_i\;=\;\frac{1}{2}\,(I\:+\:\epsilon\,\vni\cdot\vec{\sigma}),\label{eq:fuzzy}\ee
where $\epsilon\in (0,\,1)$. These operators correspond to the
events in the detectors: whenever the $i-th$ detector changes its
state, and irrespective of the actual state of other detectors,
the quantum state makes a jump implemented by the operator $a_i.$
Thus the $a_i$-s play the role of operators $\gab$: \be \gab
\doteq \sqrt{\kappa}\, a_i\ee whenever the states $\alpha$ and
$\beta$ differ just at the $i$-th place, otherwise $\gab = 0.$ The
coupling constant $\kappa$ is introduced here for dimensional
reasons. Note that for $\epsilon\,=\,1$ the $a_i$ are projection
operators. For $\epsilon<1$ Eq. (\ref{eq:fuzzy}) implies that a
projection valued measure corresponding to a sharp measurement has
been replaced by a fuzzy positive operator valued measure. Because
of this, as a result of a jump, not all of the old state is
forgotten. The new states depends, to some degree, on the old
state. Here EQT differs in an essential way from the naive von
Neumann's projection postulate of quantum theory. The parameter
$\epsilon$ becomes important. If $\epsilon=1$ -- the case where
$P(\n ,\epsilon)=P(\n )$ is a projection operator -- the new
state, after the jump, is always the same, it does not matter what
was the state before the jump. There is no memory of the previous
state, no `learning' is possible, no `lesson' is taken. This kind
of a `projection postulate' was rightly criticized in the physical
literature as being in contradiction with the real world events,
contradicting, for instance, the experiments when we take
photographs of elementary particles tracks. But when $\epsilon$ is
just close to the value $1$, but smaller than $1,$ the
contradiction disappears. This has been demonstrated in our cloud
chamber model \cite{jad94b,jad94c}, where particles leave tracks,
in real time, much like in real life, and that happens because the
multiplication operator by a Gaussian function (\ref{eq:gaussian})
does not kill the information about the momentum content of the
original wave function. Notice that the fuzzy projections $P(\n
,\epsilon)$ have properties similar to those of Gaussian
functions, namely
\begin{equation}
P(\n,\epsilon)^2=\frac{1+\epsilon^2}{2}P(\n,\frac{2\epsilon}{1+\epsilon^2}).
\label{eq:pn2} \end{equation} We describe now a sample path of the
process. Let us first discuss the algebraic operation that is
associated with each quantum jump. Suppose before the jump the
state of the quantum system is described by a projection operator
$P(\rr)$, $\rr$ being a unit vector on the sphere. That is,
suppose, before the detector flip, the spin `has' the direction
$\rr$. Now, suppose the detector $P(\n ,\epsilon )$ flips, and the
spin right after the flip has some other direction, $\rr^\prime$.
What is the relation between $\rr$ and $\rr^\prime$? It is easy to
see that the action of the operator $P(\n ,\epsilon )$ on a
quantum state vector is given, in terms of operators, by the
formula: \be \lambda(\epsilon,\n,\rr) P(\rr^\prime) =P(\n
,\epsilon )P(\rr )P(\n ,\epsilon ), \label{eq:lambda1} \ee where
$\lambda(\epsilon,\n,\rr)$ is a positive number. A simple
(although somewhat lengthy) matrix computation leads to the
following result:\footnote{The formula (\ref{eq:rprime}) is
similar to the formula for a Lorentz boost in direction $\n$ with
velocity $\beta=2\epsilon/(1+\epsilon^2).$ More information about
this analogy can be found in Ref. \cite{jad03a}.}

\be \lambda(\epsilon,\n,\rr)=\frac{1+\epsilon^2+2\epsilon
(\n\cdot\rr )}{4} \label{eq:lambda2}, \ee \be
\rr^\prime=\frac{(1-\epsilon^2)\rr+2\epsilon(1+\epsilon(\n\cdot\rr
))\n}{1+\epsilon^2+2\epsilon (\n\cdot\rr )},\ee where $(\n\cdot\rr
)$ denotes the scalar product, \be \n\cdot\rr=n_1r_1+n_2 r_2+n_3
r_3.\label{eq:rprime}\ee According to EQT the probabilities $p_i$
are computed from the formula (\ref{eq:pab}) which, in our case,
translates to \be
p_i=const\cdot\mbox{Tr}\left(P(\rr)P(\vec{n}_i,\epsilon)^2P(\rr)\right),
\ee where $const$ is the normalizing constant. Using cyclic
permutation under the trace, as well as the fact that $P(\rr
)^2=P(\rr ),$ we find, taking the trace of both sides of the
formula (\ref{eq:lambda1}), that the $p_i$ are proportional to
$\lambda(\epsilon,\vec{n}_i,\rr)$ given by (\ref{eq:lambda2}),
thus \be p_i=\frac{1+\epsilon^2+2\epsilon (\vec{n}_i\cdot
\rr)}{N(1+\epsilon^2)}. \ee Note that, owing to the fact that
$\sum_{k=1}^N \n[k]={\bf 0},$ we have $\sum_{i=1}^N p_i=1$, as it
should be. Assume that at time $t\,=\,0$ the quantum system is in
the state $\vec{r}(0)\in S^2$ (we identify here the space of pure
states of the quantum system with a two-dimensional sphere $S^2$
with radius 1). Under time evolution it evolves to the state
$\vec{r}(t)$ which is given by the rotation of $\vec{r}(0)$ with
respect to the $z$-axis. Then, at time $t_1$ a jump occurs. The
time rate of jumps is governed by a homogeneous Poisson process
with rate constant $\kappa$. When jumping $\vec{r}(t)$ moves to
$$\vec{r}_i\;=\;\frac{(1\:-\:\epsilon^2)\vec{r}(t)\:+\:2\epsilon
(1\:+\:\epsilon\vec{r}(t)\cdot\vni)\vni}{1\:+\:
\epsilon^2\:+\:2\epsilon\vec{r}(t)\cdot\vni}$$ with probability
$$p_i(\vec{r}(t))\;=\;\frac{1\:+\:
\epsilon^2\:+\:2\epsilon\vec{r}(t)\cdot\vni}{4(1\:+\:\epsilon^2)},$$
and the process starts again. The iterations lead to a
self-similar structure with a trajectory showing sensitive
dependence on the initial state, but with a clear fractal
attractor -- we may call it the `Quantum Octahedron'.
\begin{figure}[!htb]
  \begin{center}
    \leavevmode
      \includegraphics[width=11cm, keepaspectratio=true]{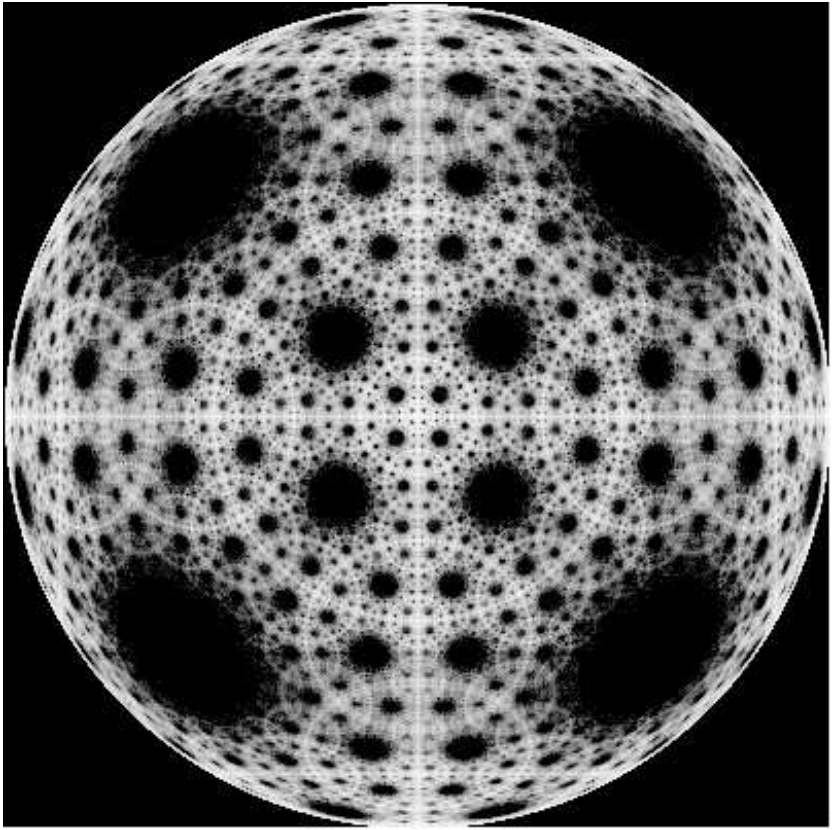}
  \end{center}
\caption{Quantum Octahedron. 100,000,000 jumps on
${\bf P}_1 ({\bC})$, $\epsilon=0.58$}
\label{octahedron}\end{figure}
This fractal figure is in the projective space ${\bf P}_1 ({\bC})$ and, though
impressive, is not what is recorded. What is recorded is a sequence of the
detector clicks, together with the times of the clicks $(n_i,t_i).$
The sequence of detector clicks looks pretty much chaotic. Here is an example:
\begin{quotation}
1,1,4,3,1,2,2,1,2,1,5,1,1,5,2,5,2,5,6,6,3,2,2,3,2,5,2,3,1,5,2,5,6,5,6,6,1,2,2,3,\\
1,3,1,4,4,5,5,5,2,5,5,6,5,6,3,5,3,2,5,4,6,5,5,6,5,4,1,4,6,6,6,3,2,6,6,5,6,5,3,2,\\
5,2,6,2,2,3,6,2,2,2,2,3,3,2,1,3,4,4,1,1,1,1,4,5,4,4,4,5,4,4,5,5,1,1,1,1,1,4,1,3,\\
2,2,5,2,2,5,5,6,4,4,3,4,5,4,5,4,4,4,6,5,2,3,2,1,1,3,3,1,5,5,5,6,3,5,6,4,5,2,2,6,\\
3,2,6,3,1,1,5,4,5,1,4,1,2,5,4,3,3,6,3,3,3,1,4,4,1,5,4,1,5,4,2,5,5,1,4,6,5,4,4,3,\\
2,3,6,5,5,5,1,5,1,1,5,5,4,6,3,1,1,1,2,2,2,1,2,2,2,5,5,5,2,2,3,6,5,2,5,5,6,4,4,4,\\
5,4,3,4,6,6,6,3,6,3,3,4,4,6,3,1,1,5,2,5,5,6,3,3,6,1,2,1,2,2,1,1,4,3,3,3,2,1,2,1,\\
4,3,4,4,4,6,1,3,6,5,1,1,1,2,2,1,1,3,4,4,3,6,2,3,1,2,3,2,5,4,4,1,6,1,1,1,1,4,6,2,\\
2,5,6,5,1,4,5,6,5,5,6,6,6,6,6,6,6,1,4,4,1,4,3,6,6,2,6,6,4,5,4,4,5,5,4,5,4,6,2,6,\\
1,5,3,4,5,6,5,6,5,5,6,3,2,3,4,5,1,4,5,4,1,4,1,5,5,6,4,1,1,3,5,1,3,6,6,6,6,6,3,2,\\
4,6,2,1,5,4,4,4,1,5,5,1,4,5,2,3,2,3,4,5,4,1,4,3,1,3,2,1,3,5,2,2,2,6,6,4,1,5
.
\end{quotation}

Any information about the initial quantum state seems to be lost rather soon,
and is probably irrecoverable from the detectors' readings due to mixing.
There is no general theory yet that would address the problem of recovering
probabilistic information about the state of the quantum system and its
dynamics from the data recorded by the classical device, except in the
limiting cases such as, for instance, when we can take Born's interpretation
limit as discussed above in section \ref{sg}.

\begin{figure}[!htb]
  \begin{center}
    \leavevmode
\includegraphics[width=11cm, keepaspectratio=true]{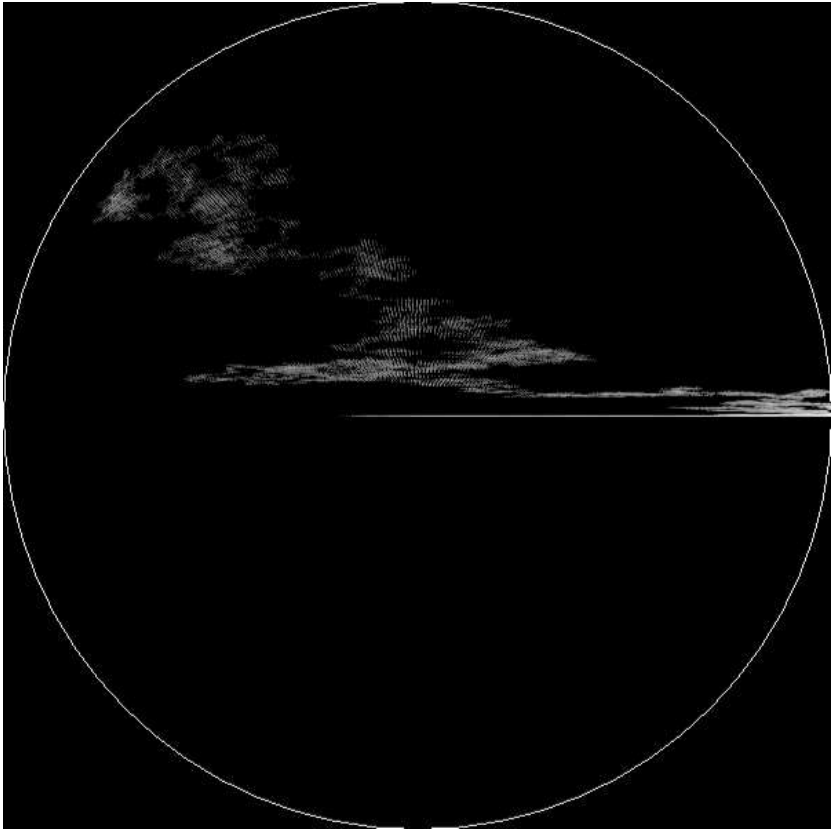}
  \end{center}
\caption{Quantum Square. 200,000 jumps on ${\bf P}_1 ({\bC})$, $\epsilon=0.0045$}
\label{square1}\end{figure}

\begin{figure}[!htb]
  \begin{center}
    \leavevmode
      \includegraphics[width=11cm, keepaspectratio=true]{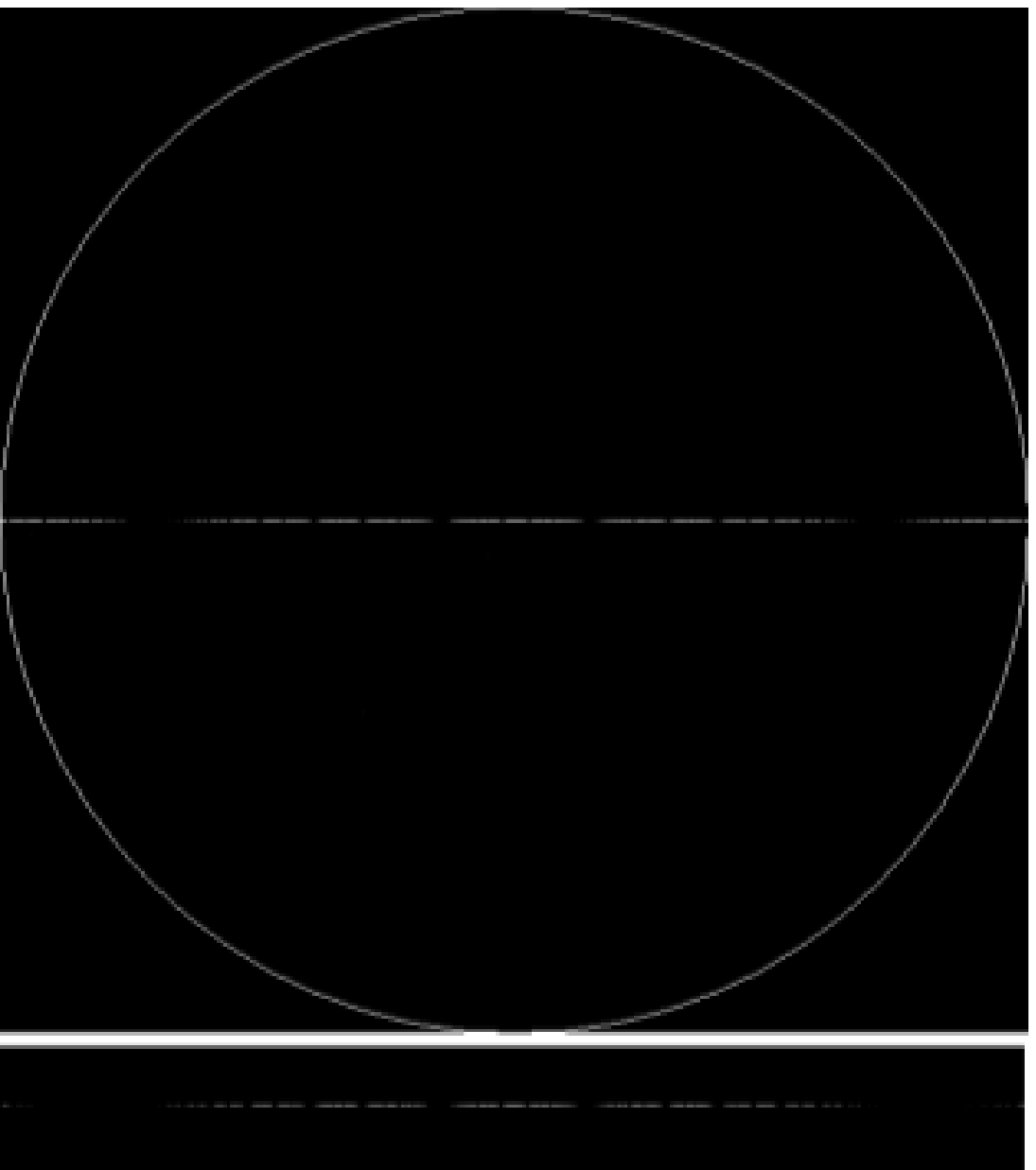}
  \end{center}
  \caption{Quantum Square. $10^6$ jumps on ${\bf P}_1 ({\bC})$, $\epsilon=0.7$.
The whole picture and $\times1000$ zoom. The square is located on
the plane perpendicular to the viewing plane. One of the four
vertices is in the center in front.} \label{square2}\end{figure}

\begin{figure}[!htb]
  \begin{center}
    \leavevmode
\includegraphics[width=11cm, keepaspectratio=true]{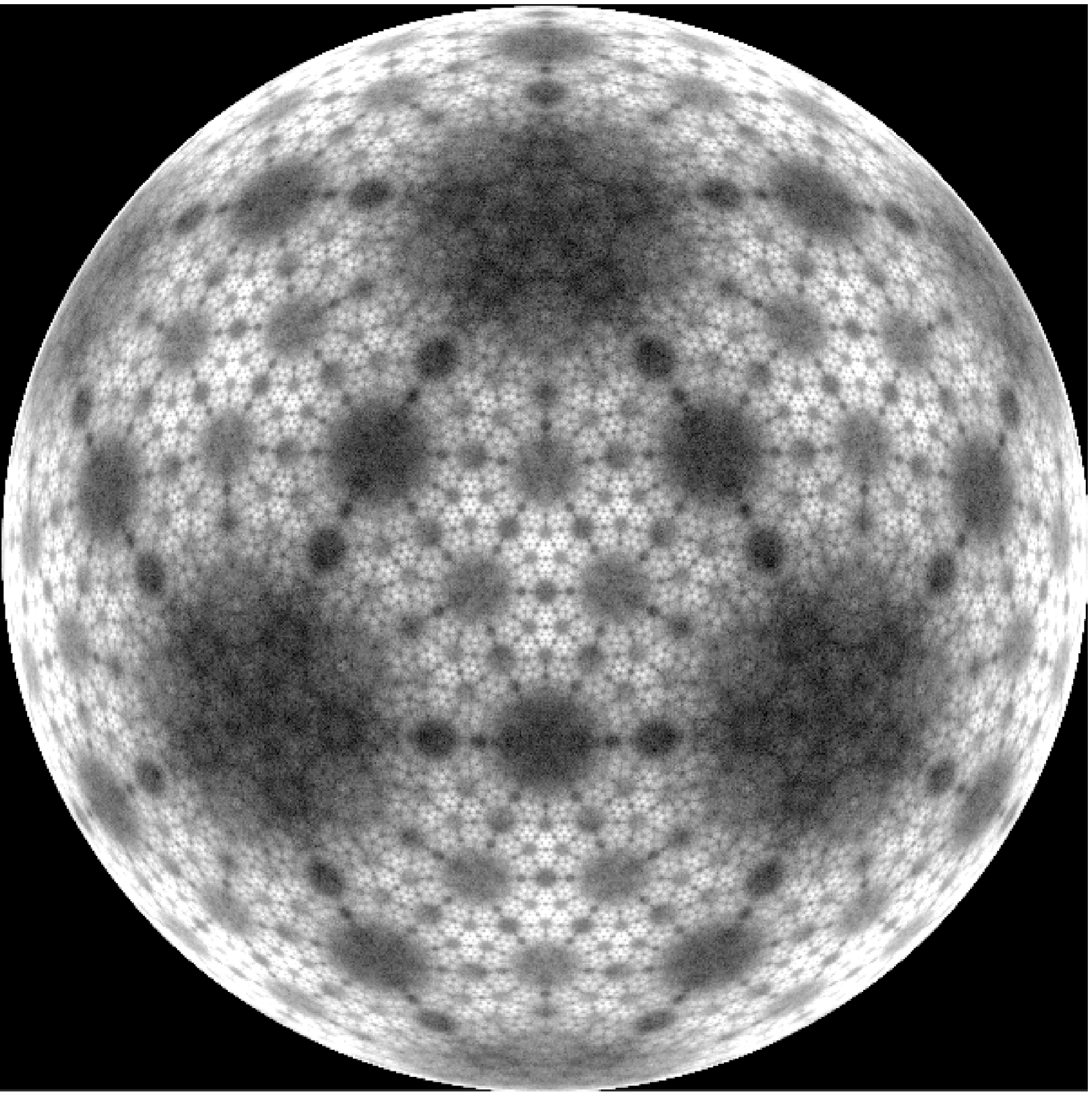}
  \end{center}
  \caption{Quantum Dodecahedron. $10^8$ jumps on ${\bf P}_1 ({\bC})$,
$\epsilon=0.65$.}\label{dodeca}
\end{figure}

Removing two vertices of the octahedron we get four points that
represent our `position-momentum' simultaneous measurement toy
model. Since the four remaining vertices are in a plane, which
intersects the sphere along a great circle, it is clear that the
attractor will be on this circle, and that the fractal pattern
will be, in this case, one-dimensional. In Figure \ref{square1} we
show the path to the attractor, starting with a randomly chosen
initial step (left upper corner). The resolution constant
$\epsilon$ had to be chosen very small, $\epsilon=0.0045$, since
otherwise the state reaches the attractor set on the circle in
just few steps. With a much higher resolution ($\epsilon=0.7$) the
Cantor set like fractal structure on the circle can be seen.
Figure \ref{square2} shows one million jumps, first the whole
picture, and then $\times1000$ zoom into the fractal attractor
set.
Figure \ref{dodeca} shows another self-similar picture,
representing quantum jumps on ${\bf P}_1 ({\bC})$ for twenty
detectors arranged at the vertices of a regular dodecahedron.

\section{Quantum Iterated Function Systems}
The EQT algorithm generating quantum jumps is similar in its
nature to a nonlinear iterated function system (IFS)
\cite{barnsley} (see also \cite{peitgen92} and references therein)
and, as such, it generically produces a chaotic dynamics for the
coupled system. IFS-s are known to produce complex geometrical
structures by repeated application of several non-commuting
affine maps. The best known example is the {\it Sierpinski
triangle}\,
\begin{figure}[!htb]
  \begin{center}
    \leavevmode
      \includegraphics[width=11cm, keepaspectratio=true]{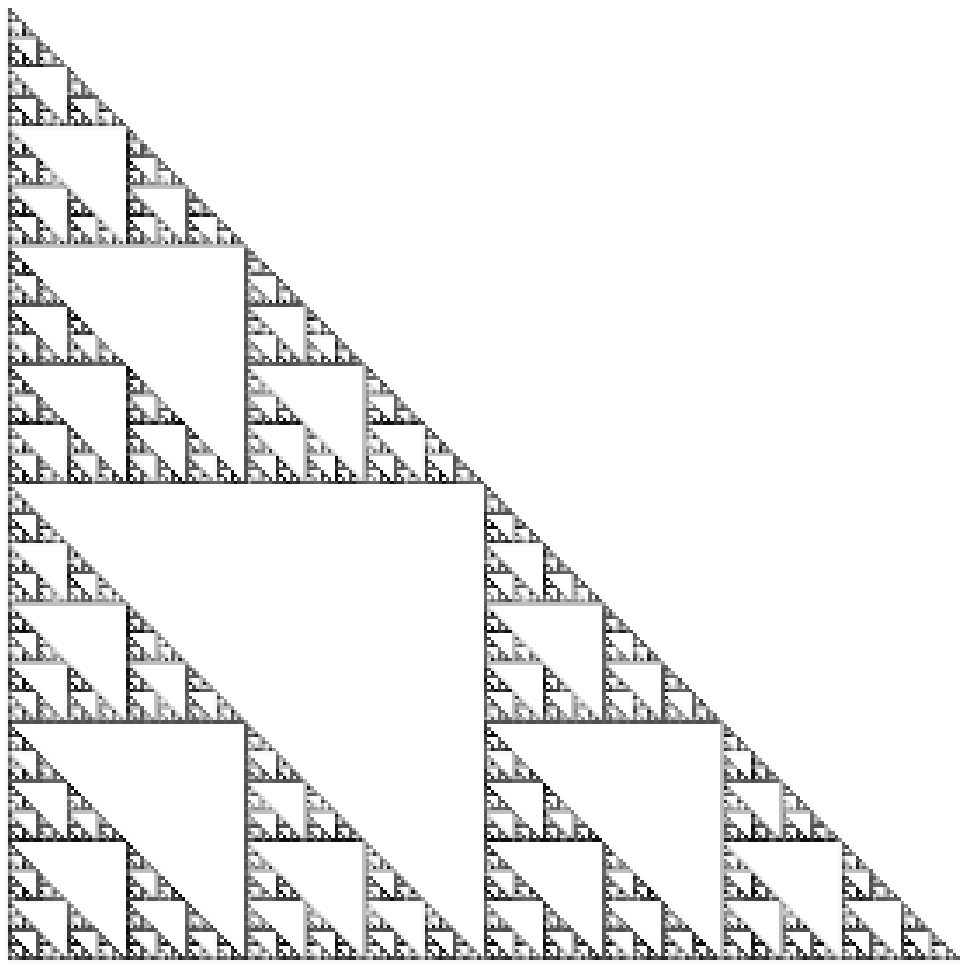}
  \end{center}
\caption{The classical fractal: a Sierpinski Triangle generated by
an Iterated Function System.}\label{sierpinski}
\end{figure}
generated by random application of $3\times 3$
matrices $A[i], i=1,2,3$ to the vector:
\begin{equation}
v_0=\left(\begin{array}{cc}x_0\\ y_0\\ 1\end{array}\right),
\end{equation}
where $A[i]$ is given by
\begin{equation}
A[i]=\left(\begin{array}{ccc}0.5&0&ax_i\cr
0&0.5&ay_i\cr0&0&1\end{array}\right)
\end{equation}
and $ax_1=1.0, ay_1=1.0, ax_2=1.0, ay_2=0.5, ax_3=0.5, ay_3=1.0.$
(Our $3\times 3$ matrices encode  affine transformations --
usually separated into a $2\times 2$ matrix and a translation
vector.) At each step one of the three transformations $A[i],
i=1,2,3$ is selected with probability $p[i]=1/3$. After each
transformation the transformed vector is plotted on the $(x,y)$
plane. Theoretical papers on IFSs usually assume that the system
is {\sl hyperbolic}\, i.e. that each transformation is a {\sl
contraction}, or, in other words, that the distances between
points get smaller and smaller. This assumption is not necessary
when the transformations are non-linear and act on a compact space
-- as is in the case of quantum fractals with which we are
dealing. In our case the probabilities assigned to the maps are
derived from quantum transition probabilities and thus  depend on
the actual point, but such generalizations of the IFS's have been
also studied (cf. \cite{peigne} and references therein). Our
algorithm generates {\em quantum fractals}\footnote{The term {\em
quantum fractals\/} has been used before by Casati et al.
\cite{casati91,casati99} in a different context.}, that is
self-similar patterns on the complex projective space of pure
states of a quantum system.

In a recent paper \cite{loslzy} {\L}ozinski, S{\l}omczynski and Zyczkowski
studied iterated function systems on the space of mixed states, when
probabilities that are associated with maps are given independently of the maps,
and under assumptions that the maps are invertible. Our maps $\gab$ do not have to be invertible, because the probability generating formula (\ref{eq:pab}) assures that probability zero is assigned to a map $\psi\mapsto \gab \psi/\Vert \gab \psi\Vert$ whenever $\Vert \gab \psi\Vert = 0.$
\section{Open Problems in EQT}
There are two kinds of open problems in EQT: those dealing with
the developing theory and its applications, and those related to
its very foundations. Let us start with questions of the first
kind. The original motivation for creating EQT was our
dissatisfaction with the rigor of Hawking's derivation of the
black hole radiation formula. The way the classical gravitation
field was coupled to a quantum field, with back reaction of the
quantum field on classical gravitation,  was, in our opinion,
highly unsatisfactory. Then there was gravity itself, where the
ten components of the metric tensor split naturally into a scalar
field $\Phi$ that defines the volume form, and the nine components
that define the conformal (i.e. light cone) structure of
space-time. A first attempt in quantizing gravity would be
quantizing $\Phi$, and leaving the conformal structure at the
classical level, yet coupled to the quantized $\Phi$ with a
back-action. Although EQT evolved and matured since its birth in
1993, so that we can deal now with infinitely many degrees of
freedom and continuous spectra, the two original motivating
problems have not yet been modeled within EQT. Also the problem of
formulating EQT strictly within the algebraic framework is still
waiting for implementation.

Concerning the foundational problems, the number one problem is
the derivation of EQT from a set of primitive and easily
understandable assumptions. There are several options here and we
would like to comment on some of them.

\subsection{Environmentally Induced Decoherence}
Following the ideas introduced by Gell-Mann, Hartle, Zurek, Zeh
and others (see \cite{gellhart93,zurek91,zurek93,giulini96,zeh01}
Blanchard and Olkiewicz \cite{blaolk03} tried to rigorously derive
the emergence of classical degrees of freedom from particular
types of quantum dynamics. The main problem with this kind of an
approach is that it uses the undefined, somewhat magical, term
`environment'. How to split the universe into `the system' and
`its environment' is never discussed. If the splitting happens
only in the brain of a physicist, the phrase `dynamically induced'
is somewhat exaggerated. The point is that the authors start with
the assumed open system dynamics and Liouville equation without
ever discussing how Schr\"odinger's dynamics can `dynamically'
deform from an automorphic to a dissipative form. The formal
operations, like splitting into a tensor product and taking a
conditional expectation value are purely mathematical and have
nothing to do with real, physical, dynamical processes. As we
mentioned in the introductory part, the very use of the term
`environment', without a rigorous definition of the term, is as
useless as the use of the term measurement, without being able to
provide its formal definition. Any attempt to derive decoherence
as a limiting procedure fails to address the problem of events
happening in finite time, with the system reacting to the events
that happen.
\subsection{Bohmian Mechanics}
Bohmian mechanics (see \cite{bohmhiley,duerr}) assumes that the
classical degrees of freedom -- positions and momenta of the
particles evolve in a modified potential, determined by the
quantum wave function. This theory needs a preferred basis, like,
for instance, the coordinate basis. But in the case of pure spin
no such preferred basis exists and there is no way in which it can
be dynamically selected in a generic case. One can try to argue
that the eigenvalue decomposition of a given density matrix
provides such a basis, but even this reasoning fails when the
density matrix eigenvalues are degenerate.
\subsection{Infrared Sectors} If we believe in quantum field theory and if
we are ready to take a lesson from it, then we must admit that one
Hilbert space is not enough, that there are inequivalent
representations of the canonical commutation relations, that there
are superselection sectors associated to different phases. In
particular there are inequivalent infrared representations
associated to massless particles (cf. \cite{roep70} and references
therein). Then classical events would be, for instance, soft
photon creation and annihilation events. That idea has been
suggested by Stapp \cite{stapp85a,stapp85b} some twenty years ago,
and was analyzed in a rigorous, algebraic framework by D. Buchholz
\cite{buch94a}.

Another possibility is that not only photons, but also long range
gravitational forces may take part in the transition from the
potential to the actual. That hypothesis has been expressed by
several authors (see e.g. contributions of F. K\'arolyh\'azy et
al., and R. Penrose in \cite{penr85}; also L. Diosi
\cite{diosi89}).

\subsection{Deformation Quantization} A large part of our understanding of
quantum theory comes through the idea of `quantization'. For
instance, we take a classical Hamiltonian system on a phase space
$M$, typically $M\approx\bR^{2n}$, and we deform the product $fg$
on the space of $C^{\infty}$ functions on $M$ into a new,
non-commutative product $f\star_\lambda g$, so that we recover the
classical structure for $\lambda\rightarrow 0$ (see
\cite{bffls1,bffls2} for a review). It is in this way that we are
able to interpret algebraic objects of quantum theory, by relating
them to the well understood objects of classical Hamiltonian
dynamics. Flato and Sternheimer suggested \cite{fs97} that EQT may
be, perhaps, derived in a similar way, via deformation
quantization of a classical dissipative (and thus non-Hamiltonian)
non-Hamiltonian structure, so that the transition operators $\gab$
of EQT can be traced back to well understood classical objects.
Deformation quantizations of generalizations of standard Poisson
structures and of Hamiltonian dynamics has indeed been developed,
mainly with applications based on Nambu mechanics
\cite{df97,fds97}, and quantization of classical dissipative
structures has been studied via generalized canonical quantization
\cite{enz94,tarasov01a,tarasov01b}. Yet, until now, the program of
quantizing only a part of the system, while the other part remains
classical, and relating the result to the formal scheme of EQT
remains open.
\subsection{Natural Mathematical Constructions} Temporal evolution of a
non-dissipative quantum system is described by a one-parameter
group of automorphisms of its algebra of observables. It was a
surprising discovery when Tomita-Takesaki theory allowed us to
naturally associate such a group with each faithful normal state
(or, more generally, weight) of the algebra. Connes and Rowelli
\cite{conrov94} speculated that the modular group of automorphisms
of the equilibrium thermal state of the universe provides a
quantum dynamics at a fundamental level, a dynamics that defines,
by itself the very `rate of flow of time'. It is quite possible
that by a generalization of the Tomita-Takesaki scheme natural
semigroups of completely positive maps can be associated to
certain states of von Neumann algebras. If so, then natural
examples of EQT dynamics can be produced via pure algebraic means.
Some of these example may have a physical interpretation and
application to fundamental structures of physics.
\subsection{Concluding Remarks}
One of the reviewers of the early version of this paper wrote:
\begin{quotation}
``...it is not clear why this approach is more well-founded than
ordinary non-relativistic quantum mechanics, nor does the author
take account of the fact that most shortcomings of that theory
disappear in relativistic quantum field theory.''
\end{quotation}
In my view this is a typical misunderstanding, shared by many of
those physicists who work on difficult technical problems of
relativistic quantum fields, and do not follow the discussion
about the foundations of quantum theory and its philosophical and
interpretational problems. Relativistic quantum field theory not
only is of no help in resolving the measurement problem, but,
instead, makes it even more profound. R. D. Sorkin discussed some
the issues involved in \cite{sorkin}, where he argues that the
standard Hilbert space approach leaves relativistic quantum fields
``with no definite measurement theory, removing whatever
advantages it may have seemed to possess vis a vis the
sum-over-histories approach, and reinforcing the view that a
sum-over-histories framework is the most promising one for quantum
gravity.'' Algebraic quantum field theory is not better in this
respect, as measurements are there never defined, and no event
ever happens in a finite time. In \cite{blaja98a} we
wrote:\begin{quotation} Meaningless infinities of relativistic
quantum field theory tell us that something is seriously wrong
with our theoretical assumptions. In our opinion, the value of a
theory consists not in that it can explain the technique
 by which the fabric is woven on the loom of Nature, but that it can explain the patterns
 of the weaving, the Weaver and perhaps the motivations behind the weaving.

Facts cannot be understood by being crafted into a summary or a formula -
they can only be understood by being explained.  And, understanding is not
the same as `knowing.' Quantum Theory, as any other theory, has a finite
 region of validity - when
attempts are made to apply it beyond these limits - we get either
nonsense or no answer at all. Quantum theory, in its orthodox
version, cannot even be applied to an individual system - like the
Universe we live in and experience. We want to discover `why' in
addition to `what' regarding the order of the universe in which we
find ourselves.  We wish to discover why `this' MUST be so, rather
than `that;' why Nature does what she does and how.  We want to
uncover and understand the Laws of Nature, not just the `rules of
thumb'.''
\end{quotation}
A typical application of EQT would be, for instance, a dynamical
phase transition in the early universe. Recently a hypothesis has
been advanced that the universe should be described by a KMS state
at the Planck scale\footnote{KMS, or Kubo--Martin--Schwinger,
conditions are algebraic conditions, derived from the classical
Gibbs ensemble, that characterize a thermodynamic equilibrium
state of an infinite quantum system -- cf \cite[Ch. 5.1]{emch}}
(cf. \cite{ibog02} and references therein) and that there is a
signature fluctuation at this scale. Such a phenomenon can not be
described within the standard quantum theory, as it involves
`events' and it applies to `time itself'. Moreover, KMS condition
on a state implies automatically stationarity, therefore the
universe would always persist in its original thermodynamic
equilibrium. The only way out of an equilibrium is via a quantum
jump, using a mechanism analogous to the one described above. Such
a jump, or a sequence of jumps, could lead not only to a new
phase, but also to self-similarity that is nowadays being observed
in the Universe \cite{dalpino95}.

In EQT all the probablistic interpretations of quantum theory are
derived from the dynamics. In particular, it makes no sense to ask
the question `what would be a distribution of observed values of
an observable' without adding the appropriate terms to the
evolution equation. It is because of the dynamical treatment of
the measurement that leads to events and information transfer in
finite time, with a feedback, that EQT allows us to answer more
questions and to analyze experimental situations that the Standard
Quantum Theory seems to exclude from its consideration -- like the
simultaneous measurement of several noncommuting observables. In
this case, as has been explained in this paper, measurement
results exhibit chaotic and fractal behaviour.

\section{Acknowledgments} The author thanks QFS, Inc for support,
and to Eric Leichtnam and
Oscar Garcia-Prada for invitation to the workshop G\'eom\'etrie et
Physique, CIRM, Marseille. Thanks are also due to Palle Jorgensen
for his encouragement and helpful comments, and to my wife Laura
for reading the manuscript.


\begin{thebibliography}{99}
\bibitem{Neu} J.~von Neumann,
\newblock {\it Mathematische Grundlagen der Quantenmechanik}
        (Springer, Berlin, 1932).

\bibitem{jad94c} A.~Jadczyk,
\newblock On Quantum Jumps, Events and Spontaneous
Localization Models,
\newblock Found. Phys. {\bf 25}, 743--762  1995)

\bibitem{bell89} J.~Bell,
\newblock Towards an exact quantum mechanics,
\newblock in S.~Deser and R.~J.~Finkelstein (eds){\it Themes in Contemporary Physics II.  Essays in
honor of Julian Schwinger's 70th birthday},  (World Scientific,
Singapore 1989)

\bibitem{bell90}
J.~Bell,
\newblock Against measurement,
\newblock in {\sl
Sixty-Two Years of Uncertainty. Historical, Philosophical and
Physical Inquiries into the Foundations of Quantum Mechanics},
Proceedings of a NATO Advanced Study Institute, August 5-15,
Erice, Ed. Arthur I. Miller, NATO ASI Series B vol. 226 , (Plenum
Press, New York 1990)

\bibitem{bell87a} J.~Bell, On the impossible pilot wave, in \cite{bell86}

\bibitem{bell86}J.~Bell, Six possible worlds of quantum mechanics,
in {\sl Speakable and unspeakable in quantum mechanics},
(Cambridge University Press, 1987)

\bibitem{bell87b} J.~Bell, Beables for quantum theory, in \cite{bell86}

\bibitem{wheeler83}
John A.~Wheeler,
\newblock Delayed--Choice Experiments and
Bohr's Elementary Quantum Phenomenon
\newblock  in {\it Proc. Int.
Symp. Found. of Quantum Mechanics}, (Tokyo 1983), pp. 140--152

\bibitem{wheeler89}
John A.~Wheeler,
\newblock It From Bit, in {\it Proceedings 3rd
International Symposium on Foundations of Quantum Mechanics},
(Tokyo, 1989)

\bibitem{ingarden89} R.~Ingarden and M.~Grabowski, {\it Mechanika Kwantowa}, in Polish, (PWN, Warszawa, 1989)

\bibitem{ozawa2001}
M.~Ozawa,
\newblock Controlling Quantum State Reduction,
\newblock Phys.\ Lett. A {\bf  282}, 336 (2001).

\bibitem{ozawa2002}
M.~Ozawa,
\newblock Position measuring interactions and the
Heisenberg uncertainty principle,
\newblock Phys.\ Lett. A  {\bf  299}, 1 (2002).

\bibitem{ozawa2003a}
M.~Ozawa,
\newblock Universally valid reformulation of the Heisenberg uncertainty
principle on noise and disturbance in measurement,
\newblock Phys.\ Rev. A  {\bf  67}, 042105 (2003).

\bibitem{ozawa2003b}
M.~Ozawa,
\newblock Uncertainty Relations for Joint Measurements of non-commuting Observables,
\newblock eprint archive { \em  quant-ph/0310070}

\bibitem{ozawa2003c}
Ozawa, M.:
\newblock Uncertainty principle for conservative measurement and computing,
\newblock eprint archive {\em  quant-ph/0310071}

\bibitem{Rob29}
H.~P.~Robertson,
\newblock The uncertainty principle
\newblock Phys.\ Rev.  {\bf 34}, 163-164 (1929).

\bibitem{Ken27}
E.~H.~Kennard,
\newblock Zur Quantenmechanik einfacher Bewegungstypen
\newblock Z. Phys.  {\bf 44}, 326-352 (1927)

\bibitem{Bal70}
L.~E.~Ballentine,
\newblock The Statistical Interpretation of Quantum
Mechanics.
\newblock Rev.\ Mod.\ Phys. {\bf 42}, 358-381 (1970).

\bibitem{Per93}
A.~Peres,
\newblock {\it Quantum Theory: Concepts and Methods}
\newblock (Kluwer Academic, Dordrecht, 1993)

\bibitem{blaja93c} Ph.~Blanchard,  and A.~Jadczyk,
\newblock How and When Quantum Phenomena Become Real,
\newblock in Z.  Haba et all. (eds) {\it Proc. Third Max Born Symp., Stochasticity
and Quantum Chaos}, Sobotka 1993,  pp. 13--31,  (Kluwer Publ.
1994)

\bibitem{olk97} R.~Olkiewicz,
\newblock Some mathematical problems related to
classical-quantum interactions,
\newblock Rev. Math. Phys. {\bf 9}
(1997), 719

\bibitem{blaolk99a}  Ph.~Blanchard, R.~Olkiewicz,
\newblock Interacting quantum and
classical continuous systems I. The piecewise deterministic
dynamics,
\newblock J. Stat. Phys. {\bf 94}, 913  (1999)

\bibitem{blaolk99b}  Ph.~Blanchard, R.~Olkiewicz,
\newblock Interacting quantum and
classical continuous systems II. Asymptotic behavior of the
quantum system,
\newblock J. Stat. Phys. {\bf 94}, 933  (1999)

\bibitem{olk99} R.~Olkiewicz,
\newblock Dynamical semigroups for interacting quantum
and classical systems,
\newblock J. Math. Phys. {\bf 40}, 1300  (1999)

\bibitem{alicki87} R.~Alicki and K.~Lendi, \newblock {\it Quantum Dynamical Semigroups and
Applications}, \newblock Lect. Notes Phys. {\bf 286}, 1987

\bibitem{alicki02} R.~Alicki, \newblock {\it Invitation to quantum dynamical semigroups},
\newblock Lecture given at the 38 Winter School of Theoretical Physics, Ladek, Poland, Feb. 2002,
\newblock eprint no arXiv:quant-ph/0205188

\bibitem{emch} G.~G.~Emch, \newblock {\em Algebraic methods in
Statistical Mechanics and Quantum Field Theory},
\newblock (Wiley - Interscience, New York 1972)

\bibitem{landsman91} N.~P.~Landsman,
\newblock Algebraic Theory of Superselection Sectors and the Measurement
Problem in Quantum Mechanics, \newblock Int. J. Mod. Phys. A {\bf
6} (1991), 5349-5371

\bibitem{davis93}  M.~H.~Davis, \newblock {\em Markov models and optimization},
\newblock Monographs on Statistics and Applied Probability,  (Chapman and Hall,
London 1993)

\bibitem{blajaol99a} Ph.~Blanchard, A.~Jadczyk and R.~Olkiewicz,
\newblock Completely Mixing Quantum Open Systems and Quantum Fractals,
\newblock Physica D: Nonlinear Phenomena, {\bf 148} (3-4),
227--241 (2001)

\bibitem{chr} E.~Christensen and D.~Evans,
\newblock Cohomology of
operator algebras and quantum dynamical semigroups,
\newblock J.
London. Math.  Soc.   {\bf 20}, 358--368 (1978)

\bibitem{koss} V.~Gorini, A.~Kossakowski and E.~C.~G.~Sudarshan,
\newblock Completely positive dynamical semigroups of N--level
systems,
\newblock J.  Math.  Phys. {\bf 17}, 821--825 (1976)

\bibitem{lin} G.~Lindblad,
\newblock On the Generators of Quantum Mechanical
Semigroups,
\newblock Comm. Math. Phys. {\bf 48}, 119--130 (1976)

\bibitem{arv} W.~B.~Arveson, Subalgebras of $C^{\star}$--algebras,
Acta Math. {\bf 123}, 141--224 (1969)

\bibitem{stinespring} W.~F.~Stinespring,
\newblock Positive functions on $C^*$-algebras,
\newblock Proc. Amer. Math. Soc. {\bf 6}, 211-216  (1955)

\bibitem{jakol95} A.~Jadczyk, G.~Kondrat and R.~Olkiewicz,
\newblock On
uniqueness of the jump process in quantum measurement theory,
\newblock
J. Phys. A\, {\bf 30}, 1-18  (1996)

\bibitem{blaja95f} Ph.~Blanchard,  and A.~Jadczyk,
\newblock Time of Events in Quantum
Theory,
\newblock Helv. Phys. Acta {\bf 69}, 613--635  (1996) , also
available as quant-ph/9602010

\bibitem{muga} J.~G.~Muga, I.~L.~Egusquiza, J.~A.~Damborenea, and F.~Delgado,
\newblock Bounds and enhancements for the negative scattering-time delays,
\newblock Phys. Rev. A {\bf 66},  042115  (2002)

\bibitem{palao} J.~P.~Palao, J.~G.~Muga and A.~Jadczyk,
\newblock Barrier traversal times using a phenomenological track formation model,
\newblock Phys. Lett. A {\bf 233}, 227-232  (1997)

\bibitem{rus02a} A.~Ruschhaupt, \newblock A Relativistic Extension of Event-Enhanced Quantum Theory,
\newblock  J. Phys. A: Math. Gen. {\bf 35}, 9227-9243  (2002)

\bibitem{rus02b} A.~Ruschhaupt, \newblock Relativistic Time-of-Arrival and Traversal Time,
\newblock  J. Phys. A: Math. Gen. {\bf 35}, 10429-10443  (2002)

\bibitem{peres00b} A.~Peres,\newblock Classical interventions in quantum systems,
\newblock Phys. Rev. A {\bf 61}, 022117 (2000)

\bibitem{jad94b} A.~Jadczyk, Particle Tracks, Events and Quantum
Theory,  Progr.~Theor.~Phys. {\bf 93}, 631--646  (1995)\\ (A model
of particle tracks using a continuous medium of detectors. Gives
GRW `spontaneous localization' model as a particular case.)

\bibitem{blaja93a} Ph.~Blanchard and A.~Jadczyk, On the
Interaction Between Classical and Quantum Systems, Phys. Lett.
A{\bf 175}, 157-164  (1993) (The very first paper, describing the
method of coupling of a quantum and of a classical system via
dynamical semigroup. At that time we didn't yet know about
piecewise deterministic processes.)

\bibitem{jad94a}  A.~Jadczyk, Topics in Quantum Dynamics,
in Proc.~First Caribb. School of Math.~and Theor.~Phys.,
Saint--Francois--Guadeloupe 1993, {\it Infinite Dimensional
Geometry, Noncommutative Geometry, Operator Algebras and
Fundamental Interactions}, ed. R.Coquereaux et al., (World
Scientific, Singapore 1995) ( A review paper. Describes the PDP
process. States the problem: ``How to determine state of an
individual quantum system?'' Describes a process on $S^2$ which
later was adapted for generation of quantum fractals.)

\bibitem{popper} K.~Popper,
\newblock {\it Unended Quest. An intellectual Autobiography},
(Routledge, London 1993)

\bibitem{putnam81} H.~Putnam,
\newblock Quantum Mechanics and the Observer,
\newblock Erkenntnis {\bf 16}, 193 (1981)

\bibitem{landsman94} N.~P.~Landsman, Observation and superselection in quantum mechanics,
Stud. Hist. Phil. Mod. Phys. {\bf 26}, 45-73 (1995)

\bibitem{mackey} G.~W.~Mackey, {\it Unitary Group Representations in Physics, Probability, and Number Theory}, (Benjamin, New York 1978)

\bibitem{varadarajan} V.~S.~Varadarajan, {\it Geometry of Quantum Theory}, (Springer 1985)

\bibitem{jad03a}A.~Jadczyk and R. {\"O}berg,
\newblock Quantum Jumps, EEQT and the Five Platonic Fractals,
\newblock {\it Preprint}\ http://arXiv.org/abs/quant-ph/0204056

\bibitem{barnsley} M.~F.~Barnsley, {\it Fractals everywhere},
(Academic Press,  San Diego 1988)

\bibitem{peitgen92} H-O.~Peitgen, J.~Hartmut and D.~Saupe,
\newblock {\em
Chaos and Fractals. New Frontiers of Science}, (Springer, New
York, 1992)


\bibitem{peigne} Marc Peign\'e,
\newblock {\em Iterated Function Systems and
spectral decomposition of the associated Markov operator},
\newblock Preprint U.R.A. C.N.R.S. 305, Pr\'epublication 93 -24, Novembre
1993

\bibitem{casati91} G.~Casati,
\newblock Quantum mechanics and chaos,
\newblock in:
M.~J.~Gianoni, A.~Voros and J.~Zinn-Justin (eds), {\em Chaos and
Quantum Physics}, (North Holland, 1991)

\bibitem{casati99} G.~Casati, G.~Maspero and D.~L.~Shepelyansky,
\newblock Quantum fractal eigenstates,
\newblock Physica D {\bf 131} (1999),
311-316

\bibitem{loslzy} A.~Lozinski, K.~Zyczkowski and W.~Slomczynski,
\newblock Quantum Iterated Function Systems,
\newblock  Phys. Rev. E  {\bf 68} (2003), 046110
\newblock Preprint {\it quant-ph/0210029}

\bibitem{gellhart93} M.~Gell-Mann and J.~B.~ Hartle,
\newblock Classical Equations for
Quantum Systems,
\newblock Phys. Rev. D {\bf 47}, 3345 (1993),

\bibitem{giulini96} D.~Giulini et al. (Eds.),
\newblock {\it Decoherence and The
Appearance of a Classical World in Quantum Theory}, (Springer,
Berlin, 1996)

\bibitem{zeh01}H.~D.~Zeh,
\newblock {\it The Physical Basis of The Direction of Time},
4$^\mathrm{th}$ ed., (Springer, Berlin, 2001)

\bibitem{zurek91} W.~H.~Zurek,
\newblock Decoherence and the transition from quantum
to classical,
\newblock Physics Today {\bf 44}, 36 (1991).

\bibitem{zurek93} W.~H.~Zurek,
\newblock Preferred states, predictability, classicality
and the environment-induced decoherence,
\newblock Progress of
Theoretical Physics {\bf 89}, 281-312 (1993).

\bibitem{blaolk03} Ph.~Blanchard, R.~Olkiewicz,
\newblock Decoherence Induced Transition from Quantum to Classical Dynamics,
\newblock Rev. Math. Phys. {\bf 15}, 217-243 (2003)

\bibitem{duerr}  K.~Berndl, M.~Daumer, D.~Dürr, S.~Goldstein, N.~ Zanghi,
\newblock A Survey on Bohmian Mechanics,
\newblock Nuovo Cim. {\bf B110} (1995)
737-750

\bibitem{bohmhiley} D.~Bohm and B.~J.~Hiley,
\newblock {\em The Undivided Universe},
(Routledge, London, 1993)

\bibitem{roep70} G.~Roepstorf,
\newblock Coherent Photon States and Spectral
Condition,
\newblock Commun. Math. Phys {\bf 19} 301--314 (1970)

\bibitem{stapp85a} H.~P.~Stapp,
\newblock On the Unification of Quantum Theory
and Classical Physics,
\newblock in Lahti, P., and Mittelstaedt, P. (eds), {\em
Symposium on The Foundations of Modern Physics},  (World
Scientific, 1985)

\bibitem{stapp85b} H.~P.~Stapp,
\newblock Einstein Time and Process Time,
\newblock in  Griffin, D.R. (ed.): {\em Physics and the Ultimate
Significance of Time}, (State University of New York Press, 1986)

\bibitem{buch94a} D.~Buchholz,
\newblock Gauss' law and the infraparticle
problem,
\newblock Phys. Lett. {\bf B 174} 331--334 (1986)

\bibitem{penr85} R.~Penrose
and C.~J.~Isham,(eds)
\newblock {\em Quantum Concepts in Space and Time},
(Clarendon Press, Oxford 1986)

\bibitem{diosi89} L.~Diosi,
\newblock Models for universal reduction of
macroscopic quantum fluctuations,
\newblock Phys. Rev. A {\bf  40},1165--1174 (1989)

\bibitem{bffls1}  F.~Bayen, M.~Flato, c.~Fronsdal, A.~Lichnerowicz and
D.~Sternheimer,
\newblock Deformation Theory and Quantization. I.
Deformation of Symplectic Structures,
\newblock Ann. Phys. {\bf 111}, 61-110 (1978)

\bibitem{bffls2} F.~Bayen, M.~Flato, c.~Fronsdal, A.~Lichnerowicz and
D.~Sternheimer,
\newblock Deformation Theory and Quantization. II. Physical
Applications, Ann. Phys. {\bf 111}, 111-151  (1978).

\bibitem{fs97} M.~Flato and D.~Sternheimer,
\newblock Private communication

\bibitem{df97} G.~Dito and M.~Flato,
\newblock Generalized Abelian Deformations:
Application to Nambu Mechanics,
\newblock Lett.Math.Phys. {\bf 39}, 107-125 (1997).

\bibitem{fds97} M.~Flato, G.~Dito and D.~Sternheimer,
\newblock Nambu Mechanics,
$N$-ary Operations and their Quantization,
\newblock in D.~Sternheimer, J.~Rawnsley and S.~Gutt (eds), Proceedings of the Ascona Meeting,
{\it Deformation Theory and Symplectic Geometry}, June 1996, {\sl
Mathematical Physics Studies}, vol {\bf 20} (1997) (Kluwer, 1997).

\bibitem{enz94} C.~P.~Enz,
\newblock Hamiltonian Description and Quantization of
Dissipative Systems,
\newblock Found. Phys. {\bf 24}, 1281-1292  (1994).

\bibitem{tarasov01a}V.~E.~Tarasov,
\newblock Quantization of non-Hamiltonian Systems,
\newblock Theoretical Physics {\bf 2.}, 150-160  (2001) .

\bibitem{tarasov01b}V.~E.~Tarasov,
\newblock Quantization of non-Hamiltonian and
Dissipative Systems,
\newblock Physics Letters {\bf A 288}. No.3/4.
(2001) pp.173-182.

\bibitem{conrov94} A.~Connes, C.~Rovelli,
\newblock Von Neumann Algebra Automorphisms
and Time-Thermodynamics Relation in General Covariant Quantum
Theories,
\newblock Class.Quant.Grav. {\bf 11}, 2899-2918  (1994).

\bibitem{sorkin} R.~D.~Sorkin,
\newblock Impossible Measurements on Quantum Fields,
\newblock in  {\it Directions in General Relativity: Proceedings of the 1993 International Symposium, Maryland, Vol. 2}: Papers in honor of Dieter Brill, pages 293-305, Bei-Lok Hu and T.A. Jacobson editors, (Cambridge University Press, 1993)

\bibitem{blaja98a} Ph.~Blanchard,  and A.~Jadczyk,
\newblock A Way Out of the Quantum Trap,
\newblock in {it Open Systems and Measurement in
Relativistic Quantum Theory}\, , H.~P.~Breuer and F.~Petruccione
(Eds.), (Lecture Notes in Physics, Springer-Verlag, 1999)\\ (A
review paper. With FAQ on EQT.) {\it Preprint} quant-ph/9812081

\bibitem{ibog02} I.~Bogdanoff, \newblock The KMS State of Spacetime at the Planck Scale,
\newblock Chin. J. Phys. {\bf 40}, 149--158 (2002).

\bibitem{dalpino95}E.~M.~de Gouveia Dal Pino, A.~Hetem, J.E.~Horvath, C.~A.~W.~de Souza, T.~Villela, J.~C.~N.~de Araujo,
Evidence for a Very Large-Scale Fractal Structure in the Universe
from Cobe Measurements, Astrophysical Journal Lett. {\bf 442},
L-45-L48 (1995).


\end{thebibliography}
\end{document}